# Reducing bias in the analysis of solution-state NMR data with dynamics detectors

Albert A. Smith[1,2], Matthias Ernst[1*], Beat H. Meier[1*], Fabien Ferrage[3*]

[1]*ETH Zurich, Physical Chemistry, Vladimir-Prelog-Weg 2, 8093 Zurich, Switzerland*

[2]*Present address: Universität Leipzig, Insitut für Medizinische Physik und Biophysik, Härtelstraße 16-18, 04107 Leipzig, Germany*

[3]*Laboratoire des biomolécules, LBM, Département de chimie, École normale supérieure, PSL University, Sorbonne Université, CNRS, 75005 Paris, France.*

M.E. : maer@ethz.ch

B.M. : beme@ethz.ch

F.F. : Fabien.Ferrage@ens.fr

## Abstract

Nuclear magnetic resonance (NMR) is sensitive to dynamics on a wide range of correlation times. Recently, we have shown that analysis of relaxation rates via fitting to a correlation function with a small number of exponential terms could yield a biased characterization of molecular motion in solid-state NMR, due to limited sensitivity of experimental data to certain ranges of correlation times. We introduced an alternative approach based on 'detectors' in solid-state NMR, for which detector responses characterize motion for a range of correlation times, and reduce potential bias resulting from the use of simple models for the motional correlation functions. Here, we show that similar bias can occur in the analysis of solution-state NMR relaxation data. We have thus adapted the detector approach to solution-state NMR, specifically separating overall tumbling motion from internal motions and accounting for contributions of chemical exchange to transverse relaxation. We demonstrate that internal protein motions can be described with detectors when the overall motion and the internal motions are statistically independent. We illustrate the detector analysis on ubiquitin with typical relaxation data sets recorded at a single or at multiple high magnetic fields, and compare with results of model-free analysis. We also compare our methodology to LeMaster's method of dynamics analysis.

## I. Introduction

Nuclear magnetic resonance (NMR) is a powerful analytical tool for the investigation of the structure and dynamics of biomolecules with atomic resolution. Biomolecular dynamics of picosecond to nanoseconds are most often characterized by NMR relaxation.[1,2] The analysis of NMR relaxation-rate constants may be based on models of internal motion,[3-5] but most investigations of picosecond-nanosecond motions rely on an approach that leaves aside assumptions about the physical nature of the motions and is thus called model-free.[6-9]

Relaxation-rate constants are linked to dynamic processes through the spectral-density function, which is the Fourier transform of the correlation function.[10,11] For typical dipole-dipole interactions, this is the correlation function for internuclear vectors, which provides direct access to molecular motions. The spectral density function is probed at the eigenfrequencies of the spin system under investigation (*e.g.*, near the Larmor frequencies), and one then assumes the correlation function of motion to consist of one or several decaying exponential terms, and attempts to fit a correlation time and amplitude for each term. When using one exponential term to describe internal motion of a molecule tumbling in solution, this is referred to as the model-free approach,[9] whereas the extended model-free approach may have two or more terms to model the internal motion.[6] 'Model-free' is sometimes also applied to solid-state relaxation analysis, although the original usage referred only to a solution-state method. In solution- and solid-state NMR, the limited sampling of the spectral-density function restricts the number of terms that may be fitted, which can be a source of bias. We have recently investigated the effect of the limited information in ensembles of relaxation rates in solid-state NMR and demonstrated that analysis with inappropriate models could result, in the worst case, in parameters of dynamics whose true values are significantly outside the confidence interval of the fitted correlation times and order parameters.[12,13]

Here, we investigate whether dynamics analysis with several internal motions in solution-state NMR is likely to suffer from similar distortions as can occur in solid-state NMR. This can be easily tested by calculating rate constants for a correlation function with several exponential terms, and then testing the fit performance when a simpler, model correlation function is used to fit the calculated rate constants. We calculated a set of longitudinal, $R_1$, and transverse, $R_2$, rate constants, as well as the dipolar cross-relaxation rate constant, $\sigma_{NH}$, for a molecule tumbling isotropically in solution, with a tri-exponential correlation function for internal motions (amplitudes, $(1-S^2)A_k$, and correlation times, $\tau_c$, in Fig. 1(a)). Such a tri-exponential correlation function of an H–N bond can result, for example, from a combination of very fast, librational motions (here assumed at 1 ps), motion of the peptide plane (320 ps), and a slower loop motion involving correlated motion of several residues (3.2 ns). Note that, in reality, we could expect such dynamics to result in a distribution of correlation times for each motion, but for simplicity we assume just three discrete correlate times. The calculated relaxation dataset was fitted with a bi-exponential correlation function for internal motions. We find excellent reproduction of the rate constants in Fig. 1(b), however, the fitted correlation times and amplitudes of the exponential terms are far from the input amplitudes and correlation times. This result indicates that such a large set of relaxation rates fails to distinguish between the simple model used in the analysis and the true, more complex model of the internal motion. Clearly, the subsequent mechanistic interpretation of results of the analysis of relaxation rates with a model that is too simple would lead to an erroneous picture of dynamics (further examples are shown in Fig. S6).

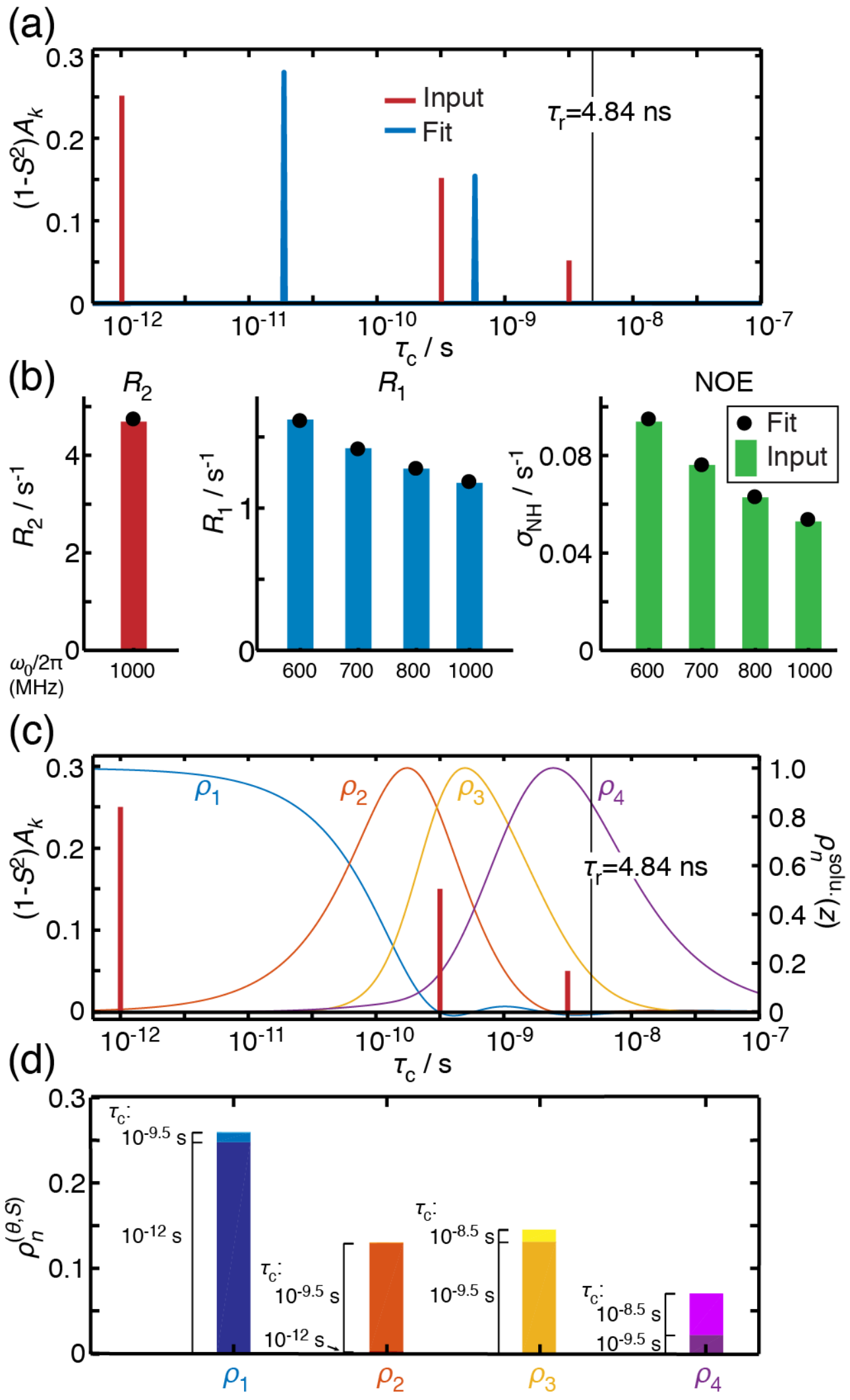

**Fig. 1.** Problematic fit behavior in solution-state NMR: Synthetic data for $^{15}$N $R_2$ at 1000 MHz, $^{15}$N $R_1$ and $^{1}$H–$^{15}$N $\sigma_{NH}$ (measured from nuclear Overhauser effects, NOE), at 600, 700, 800, and 1000 MHz for H–N backbone dynamics in solution-state NMR, for a correlation function with three correlation times. The input correlation function is shown as a function of the correlation time in (a) (red lines), with amplitudes of motion shown on the *y*-axis for a protein tumbling with = 4.84 ns. The resulting rate constants ((b), bars) are then fitted to a model correlation function having only two internal correlation times. The fitted amplitudes and correlation times are shown in (a) (blue lines), and the calculated rate constants are shown in (b) (black circles). Although a close-to-perfect fit of the rate constants is obtained, the resulting amplitudes and correlation times are far away from the input motion. Note that $R_2$ rate constants obtained at different fields contain very little independent information, so that we only show a single rate constant here (multiple $R_2$ rate constants could also easily be fit). (c) plots the sensitivities of a set of four detectors that are calculated using this data set. The amplitudes and correlation times of the input correlation function are re-plotted (red) to show the overlap of the motion and the sensitivities. (d) shows the detector responses, which give the overlap of the sensitivities with amplitudes and correlation times. Bars are separated into sections, indicating how each motion contributes to the total detector responses.

A number of approaches already address these shortcomings, but each has limitations. Spectral-density mapping, for example, determines the values of the spectral density function only at a few frequencies that determine measured relaxation rates.[14-16] This requires minimal assumptions about the complexity of motion, and so limits biasing. While the original method uses $R_1$, $\sigma_{NH}$, and $R_2$ at a single field, it is possible to exploit near-coincidence of frequencies in multi-field data sets to obtain further information.[17,18] However, because it does not retrieve the correlation times or amplitudes of motional modes, the interpretation of spectral density mapping is mostly qualitative,[19] and does not separate contributions from internal and overall (tumbling) motion. Other attempts have been made to recover information about the correlation times of motions that lead to relaxation with minimal bias. For instance, the interpretation of motions by projection on an array of correlation times (IMPACT) determines the distribution of correlation times through a simple regularization method and was applied to the analysis of relaxation in disordered proteins.[20] Similar to spectral density mapping, IMPACT does not remove the influence of tumbling. Finally, LeMaster developed an approach in which $R_1$, $\sigma_{NH}$, and $R_2$ are fitted, using fixed correlation times.[21] In this approach, LeMaster was successful in separating internal motion from tumbling, but his approach is limited to analyzing data sets recorded at a single magnetic field.

To address distortions from using an over-simplified model of the correlation function in solid-state NMR (sometimes also referred to a model-free), we have recently introduced an approach based on dynamics detectors, which are linear combinations of relaxation-rate constants, where the linear combinations are optimized to yield information about different ranges of correlation times.[12,13] A set of detectors is built for each relaxation dataset, based on the relaxation-rate constants measured, for example, at different magnetic fields. Then resulting detector sensitivities indicate what range of correlation times the set of experiments is sensitive to, and further indicate how well one may resolve different ranges of correlation times. Experimental data analysis then quantifies how much motion is in the sensitive range of each detector. More precisely, detectors yield the overlap of a sensitivity function and a distribution of correlation times of motion. For example, in Fig. 1(c) four detector sensitivities are shown, where the overlap of these sensitivities with the three correlation times and amplitudes in Fig. 1(a)/(c) results in the detector responses shown in Fig. 1(d). This approach provides quantitative information about the correlation times and amplitudes of motions with minimal assumptions about the motions.[13] Detector sensitivities clearly indicate the range of correlation times that is probed by the set of experiments and the resolution at which correlation times can be defined. By contrast, modeling the correlation function with a few decaying exponentials results in correlation times that are a function of the internal motion and of the choice of experiments.[12] As with spectral-density mapping and the IMPACT approach, detectors as previously described do not allow separation of tumbling and internal motion (Fig. 1(c)/(d) shows detectors that *do* separate tumbling and internal motion, using the methodology that we will present below).

Detectors characterize the overlap of the detector sensitivities with the distribution of motion, so that the resulting dynamics description may seem rather imprecise as compared to the seemingly well-defined correlation times and amplitudes (order parameters) resulting from modeling the correlation function as a few decaying exponentials. However, as we have shown in Fig. 1 and previously,[12] these correlation times can represent poorly defined averages of the 'true' correlation times of multiple motions. To better understand the complications brought about by this averaging process, we consider another example. We continue with the assumption of a motional model defined by three correlation times, now fixed at correlation times for fast motion, $\tau_{\text{f,in}}$ = 1 ps, for intermediate motion, $\tau_{\text{i,in}}$ = 300 ps, and for slow motion, $\tau_{\text{s,in}}$ = 3 ns. The order parameters for the two motions with shorter correlation times are also fixed, at $(1-S^2)A_{\text{f,in}} = 0.25$, $(1-S^2)A_{\text{i,in}} = 0.15$ (where $(1-S^2)A_{\text{f,in}} = (1-S^2_{\text{f,in}})$, $(1-S^2)A_{\text{i,in}} = S^2_{\text{f,in}}(1-S^2_{\text{i,in}})$). The amount of slow internal motion, corresponding to the correlation time $\tau_{\text{s,in}} = 3$ ns, is allowed to vary from $(1-S^2)A_{\text{s,in}} = 0$ to $(1-S^2)A_{\text{s,in}} = 0.21$ ( $(1-S^2)A_{\text{s,in}} = S^2_{\text{f,in}}S^2_{\text{i,in}}(1-S^2_{\text{s,in}})$ ). Relaxation-rate constants resulting from this correlation function are then fitted using a bi-exponential correlation function (using the same set of simulated relaxation data as in Fig. 1), with results in Fig. 2(a).

When $(1-S^2)A_{\text{s,in}} = 0$, the input motion has only two terms, so the parameters of the fitted model match the input model, yielding a perfect fit. When $(1-S^2)A_{\text{s,in}}$ increases, the fitted $(1-S^2)A_{\text{s,fit}}$ also increases, since it now contains contributions from both $(1-S^2)A_{\text{s,in}}$ and $(1-S^2)A_{\text{i,in}}$. Correspondingly, $\tau_{\text{s,fit}}$ also increases from contributions from the slow motion. However, we also see that there is about a 15% increase in $(1-S^2)A_{\text{f,fit}}$, and more than an order of magnitude change in $\tau_{\text{f,fit}}$. Then, we see that a very slow motion (here 3 ns) can influence a fitted parameter that corresponds to motion two orders of magnitude faster (with $\tau_{\text{f,fit}} \leq 25$ ps). This relayed influence of a slow motion on fast-motional parameters can potentially convolute the interpretation of model-free results.

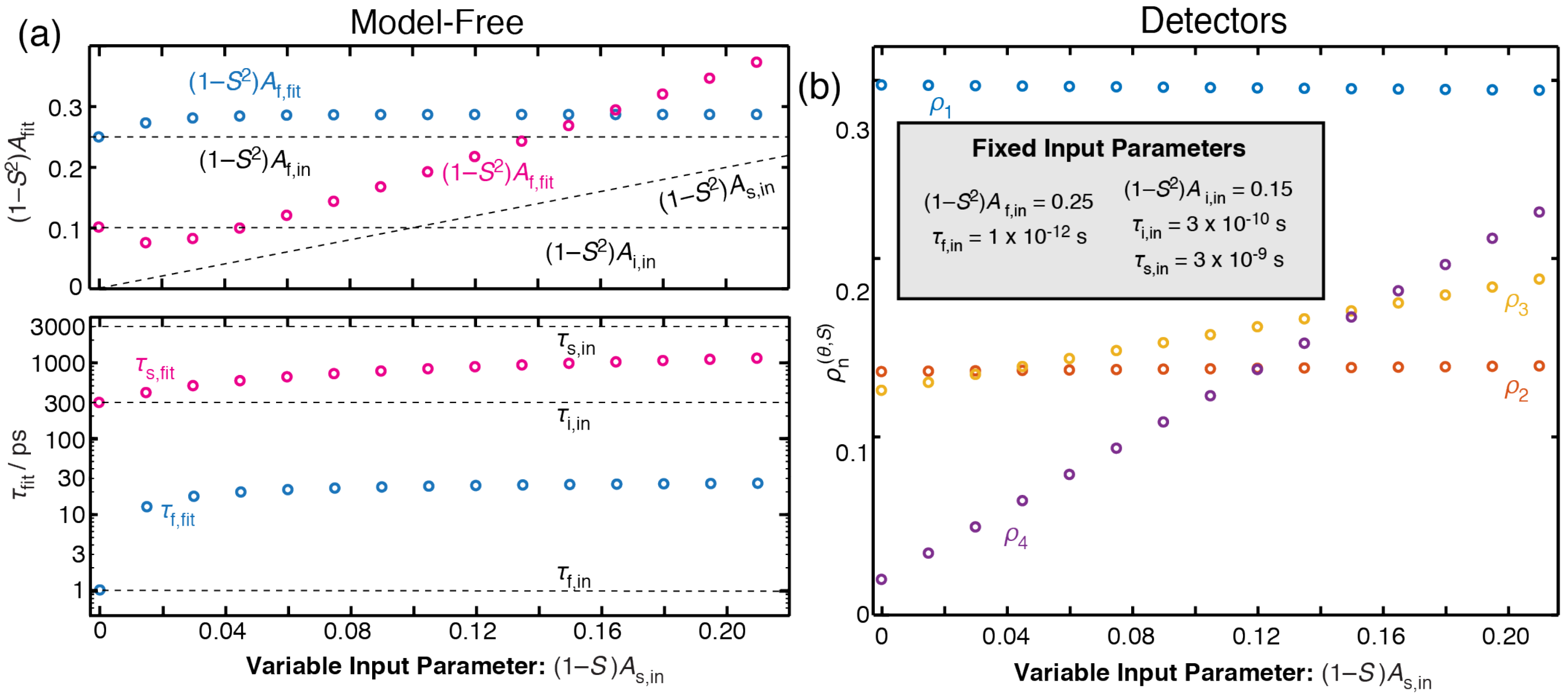

**Fig. 2.** Model-free and detector behavior as a function of motional amplitude. We assume a model with three correlation times, with all parameters but the order parameter of the slowest motion fixed ($(1-S^2)A_{\text{f,in}} = 0.25$, $\tau_{\text{f,in}} = 1\times10^{-12}$ s, $(1-S^2)A_{\text{i,in}} = 0.15$, $\tau_{\text{i,in}} = 3\times10^{-10}$ s, $\tau_{\text{s,in}} = 3\times10^{-9}$ s). (a) shows fitted amplitudes (top) and correlation times (bottom) as a function of the amplitude of the input slow motion, $(1-S^2)A_{\text{s,in}}$, resulting from fitting rate constants using a bi-exponential correlation function (same experiments as in Fig. 1). Black dotted lines show the input parameters. (b) shows detector responses for the same motions, using the detector sensitivities given in Fig. 1(c).

The effect of the change in amplitude $(1-S^2)A_{\text{s,in}}$ on the detectors analysis is more regular (Fig. 2b). Increasing the amplitude results in a strong increase in $\rho_4^{(\theta,S)}$, which is expected since the maximum of the $\rho_4$ detector sensitivity is close to 3 ns, and a smaller increase in $\rho_3^{(\theta,S)}$, since the $\rho_3$ sensitivity is nonzero at 3 ns (Fig. 1(c)). Critically, $\rho_1^{(\theta,S)}$ and $\rho_2^{(\theta,S)}$ are not visibly influenced by the slow motion, since they are only sensitive to short correlation times. An increase in the amplitude in a motion results simply in the increase in detector responses sensitive to the correlation time of that motion, with no effects on the other detectors.

Neither model-free analysis nor detectors allow us to recover a complete description of the original motion. On the other hand, given the original motion, we may directly determine detector responses (which will be precisely defined below) but we may not easily determine the model-free parameters except for special cases.[9,22] While both methods leave ambiguity in describing the original motion, detector sensitivities give a clear indication as to where these ambiguities are, via detector sensitivities. These advantages are particularly important when comparing NMR analyses to other methods, such as molecular dynamics simulations.[23]

Here, we present a modified detectors framework adapted to the analysis of solution-state relaxation, where the influence of the overall rotational diffusion (tumbling) of a macromolecule on the detector responses is removed. We have also expanded the DIFRATE software with updated methodology,[24] and analyzed typical datasets recorded at one to three static magnetic fields. We compare the results to analyses of data using the extended model-free approach, and find that we obtain a more stable, and easier to interpret description of the internal dynamics. We also compare our approach to that of LeMaster for the analysis of relaxation rate constants recorded at a single magnetic field; the methods yield very similar behavior, so that LeMaster's approach may be considered as a special case of the detectors approach.[21]

## II. Theory

### A. Background

In NMR dynamics, one often assumes that the internal motion may be described by a correlation function, $C_\mathrm{I}(t)$, consisting of one or more exponential terms,[9] so that one can write

$$C_\mathrm{I}(t) = S^2 + (1-S^2)\sum_k A_k \exp(-t/\tau_k) \quad (1)$$

Then, $(1-S^2)$ is related to the total amplitude of internal motion, and the $A_k$ give contributions from individual internal motions at effective correlation times $\tau_k$ (the $A_k$ sum to 1). In the case of solution-state NMR we usually assume separability (statistic independence)[25] of internal and overall motions leading, for isotropic tumbling, to a total correlation function of

$$\begin{aligned} C(t) &= C_\mathrm{O}(t)C_\mathrm{I}(t) \\ C_\mathrm{O}(t) &= \frac{1}{5}\exp(-t/\tau_\mathrm{r}), \end{aligned} \quad (2)$$

where $C_\mathrm{O}(t)$ is the correlation function of the overall motion, and $\tau_\mathrm{r}$ is the corresponding rotational correlation time.[9]

From $C(t)$, we obtain the spectral-density function

$$J(\omega) = 2\int_0^\infty C(t)\cos(\omega t)\,dt, \quad (3)$$

and subsequently calculate various relaxation-rate constants. In this study we will primarily concentrate on $R_1$, $R_2$, and the dipolar cross-relaxation rate constant, $\sigma_\mathrm{IS}$ (measured through nuclear Overhauser effects (NOE)).

$$R_1 = \left(\frac{\delta_{IS}}{4}\right)^2 \left(J(\omega_I - \omega_S) + 3J(\omega_I) + 6J(\omega_I + \omega_S)\right) + \frac{1}{3}\left(\omega_I \Delta\sigma_I\right)^2 J(\omega_I). \tag{4}$$

$$R_2 = \frac{1}{2}R_1 + \left(\frac{\delta^{IS}}{4}\right)\left(3J(\omega_S) + 2J(0)\right) + \frac{2}{9}\left(\omega_I \Delta\sigma_I\right)^2 J(0). \tag{5}$$

$$\sigma_{IS} = \left(\frac{\delta_{IS}}{4}\right)^2 \left(-J(\omega_I - \omega_S) + 6J(\omega_I + \omega_S)\right). \tag{6}$$

Here, $\delta_{IS}$ is the anisotropy of the dipolar coupling ($\delta_{IS} = (\mu / 2\pi)(\hbar\gamma_I\gamma_S / r^3)$) and $\omega_I\Delta\sigma_I = 3/2\delta_I$ is the difference between $\omega_I\sigma_{zz}$ and $\omega_I\sigma_{xx}$, two of the principal values of the chemical-shift anisotropy (CSA) tensor (where we assume the CSA to be axially symmetric). In this study, for $^{15}$N relaxation, these terms correspond to the $^1$H–$^{15}$N dipole-dipole, and $^{15}$N CSA interactions.

A common strategy for the determination of internal dynamics in a molecule is to measure a set of relaxation-rate constants and assume a number of exponential terms describing the internal dynamics (Eq. (1)). The correlation times ($\tau_k$) and amplitudes ($A_k$) are optimized for each exponential term such that experimental relaxation-rate constants are reproduced well. For solid-state NMR, such an approach to analysis may yield a distorted representation of the internal dynamics, if the model contains fewer exponential terms than the real motion.[12]

An alternative approach is to characterize the motion with several detector responses, which quantify the motion for a range of correlation times, defined by $\rho_n(z)$ (the detector "sensitivity"), and is unbiased by any model of the correlation function. We shortly summarize this approach here (for a detailed description, see ref [13]). Detectors are obtained via optimized linear combination of the experimental rate constants. If, for example, we take two rate constants, $R_\zeta^{(\theta,S)}$ and $R_\xi^{(\theta,S)}$, and add them together with coefficients $a$ and $b$, we can define a detector response, $\rho_n^{(\theta,S)}$, as

$$\rho_n^{(\theta,S)} = aR_\zeta^{(\theta,S)} + bR_\xi^{(\theta,S)} \tag{7}$$

We can understand why such an approach is useful, if we describe the correlation function by a distribution of correlation times of motion (henceforth referred to as the distribution of motion)

$$C(t) = \frac{1}{5}\left[S^2 + (1-S^2)\int_{-\infty}^{\infty} \theta(z)\exp(-t/(10^z \cdot 1\,\mathrm{s}))dz\right], \tag{8}$$

where (1–$S^2$) gives the total amplitude of motion, and $\theta(z)$ gives the distribution of that motion over all correlation times ($\theta(z)$ integrates to 1), where $z = \log_{10}(\tau_c / 1\,\mathrm{s})$. Then, each relaxation-rate constant is given by

$$R_\zeta^{(\theta,S)} = (1-S^2)\int_{-\infty}^{\infty} \theta(z) R_\zeta(z)\,dz \qquad (9)$$

where $R_\zeta^{(\theta,S)}$ is the rate constant for an experiment, indicated by $\zeta$, with a distribution given by $(1-S^2)\theta(z)$. $R_\zeta(z)$ is the "sensitivity" of that experiment at a given correlation time, $z$, and can be calculated from Eqs. (4)-(6), by assuming a mono-exponential correlation function with correlation time $\tau_c = 10^z \cdot 1$ s and order parameter $(1-S^2)=1$. A glossary of the terms used here is given at the beginning of the Supporting Information.

The value of $\rho_n^{(\theta,S)}$ is given by

$$\rho_n^{(\theta,S)} = (1-S^2)\int_{-\infty}^{\infty} \theta(z) \rho_n(z)\,dz \qquad (10)$$

where the sensitivity of the detector, $\rho_n(z)$, is

$$\rho_n(z) = aR_\zeta(z) + bR_\xi(z) \qquad (11)$$

One adjusts $a$ and $b$ to optimize the form of $\rho_n(z)$. This principle can be applied to large sets of experimental rate constants, so that one may design the detector sensitivities, $\rho_n(z)$, to give optimally separated ranges of correlation times. In this case, we define detection vectors, $\vec{r}_n$, which relate the experimental rate constants to the detector responses, as

$$\begin{pmatrix} \rho_1^{(\theta,S)} \\ \vdots \\ \rho_n^{(\theta,S)} \end{pmatrix} = \begin{pmatrix} [\vec{r}_1]_\zeta / \sigma(R_\zeta) & \cdots & [\vec{r}_n]_\zeta / \sigma(R_\zeta) \\ \vdots & \ddots & \vdots \\ [\vec{r}_1]_\xi / \sigma(R_\xi) & \cdots & [\vec{r}_n]_\xi / \sigma(R_\xi) \end{pmatrix}^{-1} \begin{pmatrix} R_\zeta^{(\theta,S)} / \sigma(R_\zeta) \\ \vdots \\ R_\xi^{(\theta,S)} / \sigma(R_\xi) \end{pmatrix}, \qquad (12)$$

where $[\vec{r}_j]_\zeta$ is the element of detection vector $j$, corresponding to the relaxation-rate constant denoted by $\zeta$, and the matrix power of -1 indicates a pseudo-inverse (since one typically has more experiments than detectors). $\sigma(R_\zeta)$ indicates the standard deviation for the experiment denoted by $\zeta$. Inclusion of this term re-weights the linear combination depending on data quality for each experiment and residue. It may also be omitted, but its inclusion is default in the DIFRATE software.[24] Essentially, we are fitting the measured rate constants with a sum of the detection vectors. Note that, in practice one restricts the allowed values of the detector responses, so that a linear least-squares solver may be necessary for this fit, as opposed to using a simple matrix inversion as shown here.

## B. Sensitivity to internal motion

The simple, linear relationship between the distribution of motions, $(1-S^2)\theta(z)$, and the measured rate constants, $R_\zeta^{(\theta,S)}$, as obtained in Eq. (9), is particularly useful for dynamics analysis in solid-state NMR. When no tumbling is present, the correlation function primarily describes internal motion with additional contributions from small-amplitude overall motion of the protein (such as 'rocking' in a crystal[26,27] or overall motion in a fibril[28]). By contrast, in solution-state NMR, the total correlation function is a product of the correlation function of the internal motion and the correlation function of the tumbling (Eq. (2), assuming statistical independence of the two motions). Although one may apply the detector analysis as derived for solid-state NMR directly to solution-state data, the resulting detector responses convolute information about the distribution of internal motion with information about the overall tumbling (see below). We would rather characterize only the distribution of internal motion which requires a similar relationship between the measured rate constants, $R_\zeta^{(\theta,S)}$, and the distribution of internal motion (denoted as $(1-S^2)\theta(z_i)$, where $z_i$ is the log of the internal correlation time, $z_\mathrm{i}=\log_{10}(\tau_\mathrm{i}/1\,\mathrm{s})$).

To do so, we begin with the correlation function of an interaction in a molecule undergoing isotropic molecular tumbling and internal motion described by a distribution $(1-S^2)\theta(z_\mathrm{i})$, which is given by

$$C(t)=\frac{1}{5}\exp(-t/\tau_\mathrm{r})\left[S^2+(1-S^2)\int_{-\infty}^{\infty}\theta(z_\mathrm{i})\exp(-t/(10^{z_\mathrm{i}}\cdot 1\,\mathrm{s}))dz_\mathrm{i}\right]. \tag{13}$$

In analogy to Eq. (9) this leads to a solution-state relaxation-rate constant of the form

$$R_\zeta^{(\theta,S)}=S^2R_\zeta(z_\mathrm{r})+(1-S^2)\int_{-\infty}^{\infty}\theta(z_\mathrm{i})R_\zeta\left(z_\mathrm{eff}(z_\mathrm{i})\right)dz_\mathrm{i}. \tag{14}$$

Here, $z_\mathrm{r}=\log_{10}(\tau_\mathrm{r}/1\,\mathrm{s})$, and the effective correlation time describing the combined effects of overall and internal motion is given by

$$\tau_\mathrm{eff}=\frac{\tau_\mathrm{i}\tau_\mathrm{r}}{\tau_\mathrm{i}+\tau_\mathrm{r}}. \tag{15}$$

The dependence of $\tau_\mathrm{eff}$ as a function of $\tau_\mathrm{i}$ is plotted in Fig. 3. If we take $z_\mathrm{eff}=\log_{10}(\tau_\mathrm{eff}/1\,\mathrm{s})$, we obtain

$$z_\mathrm{eff}(z)=\log_{10}\left(\frac{\tau_\mathrm{r}10^{z_i}}{\tau_\mathrm{r}+10^{z_i}\cdot 1\,\mathrm{s}}\right). \tag{16}$$

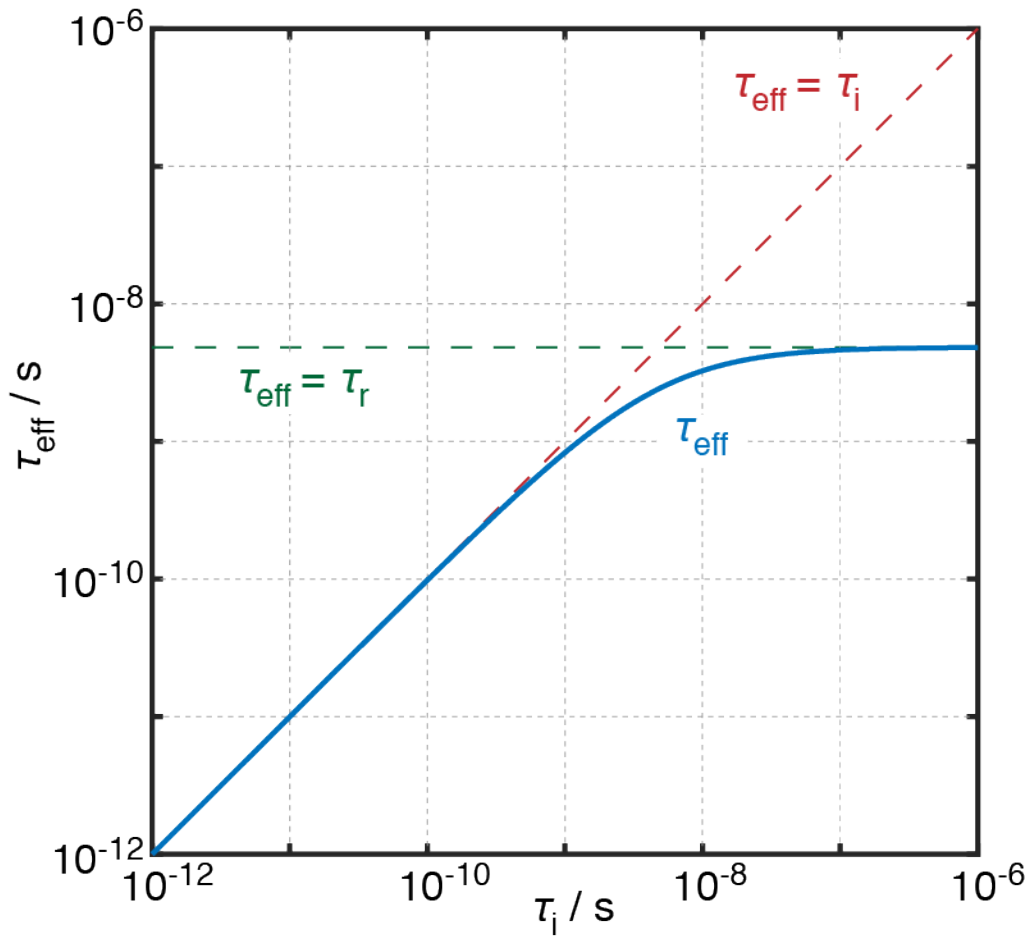

**Fig. 3.** Effective correlation time for internal motions. The effective correlation time, $\tau_{\mathrm{eff}}$, is plotted against the internal correlation time, $\tau_{\mathrm{i}}$, assuming a rotational correlation time of $\tau_{\mathrm{r}} = 4.84$ ns. We show the effective correlation time $\tau_{\mathrm{eff}}$ (solid blue line), the correlation time for internal motions, $\tau_{\mathrm{i}}$ (red dashed line), and the rotational correlation time, $\tau_{\mathrm{r}}$ (green dashed line). If $\tau_{\mathrm{i}} << \tau_{\mathrm{r}}$, then $\tau_{\mathrm{eff}} = \tau_{\mathrm{i}}$, but, as $\tau_{\mathrm{i}}$ approaches $\tau_{\mathrm{r}}$ the effective correlation time evolves asymptotically towards $\tau_{\mathrm{r}}$.

We can now rewrite the solution-state relaxation-rate constant such that the effect of overall rotational tumbling is separated from the net effects of the distribution of internal motion, $(1-S^2)\theta(z_{\mathrm{i}})$:

$$\begin{aligned} R_\zeta^{(\theta,S)} &= S^2 R_\zeta\left(z_{\mathrm{r}}\right) + (1-S^2)\int_{-\infty}^{\infty} \theta(z_{\mathrm{i}}) R_\zeta\left(z_{\mathrm{eff}}(z_{\mathrm{i}})\right) dz_{\mathrm{i}} \\ &= R_\zeta^0 + (1-S^2)\int_{-\infty}^{\infty} \theta(z_{\mathrm{i}})\left(R_\zeta\left(z_{\mathrm{eff}}(z_{\mathrm{i}})\right) - R_\zeta^0\right) dz_{\mathrm{i}} \end{aligned}, \tag{17}$$

where we have defined $R_\zeta^0 = R_\zeta(z_{\mathrm{r}})$. Then, if we define the sensitivity to internal motion as

$$R_\zeta^{\mathrm{solu.}}(z_{\mathrm{i}}) = R_\zeta(z_{\mathrm{eff}}(z_{\mathrm{i}})) - R_\zeta^0 = R_\zeta\left(\log_{10}\left(\frac{\tau_r 10^{z_{\mathrm{i}}}}{\tau_r + 10^{z_{\mathrm{i}}} \cdot 1\,\mathrm{s}}\right)\right) - R_\zeta^0, \tag{18}$$

we obtain the following formula for the relaxation-rate constant:

$$R_\zeta^{(\theta,S)} = R_\zeta^0 + (1-S^2)\int_{-\infty}^{\infty} \theta(z_{\mathrm{i}}) R_\zeta^{\mathrm{solu.}}(z_{\mathrm{i}}) dz_{\mathrm{i}}. \tag{19}$$

The resulting equation has nearly the same form as Eq. (9), with the only differences between Eq. (18) and Eq. (9) being the offset term, $R_\zeta^0$, and that we first calculate the effective correlation time from $z_{\mathrm{i}}$ and $\tau_{\mathrm{r}}$, which is then inserted into the sensitivity, as $R_\zeta(z_{\mathrm{eff}}(z_{\mathrm{i}}))$. Note that in this study, we assume isotropic tumbling throughout. In principle, one may also introduce a more complex form of the correlation function of the tumbling in Eq. (13), as would result from anisotropic tumbling. This will result in different experimental sensitivities to internal motion, $R_{\zeta}^{\text{solu.}}(z_{\text{i}})$, depending on the relative orientation of the rotational diffusion tensor and the corresponding bond. While this will make the optimization of detector sensitivities more complicated, variations in the experimental sensitivities will not prohibit the application of detectors unless the anisotropy is extreme. It is possible to generate very similar detector sensitivities for all orientations for the range $0.2 \leq D_{\parallel} / D_{\perp} \leq 5$, where variation in experimental sensitivity will require re-optimization of detector sensitivities for each bond orientation, although this step can be automated. For larger anisotropies, detector sensitivities for different residues may be significantly different, so that the detector responses themselves should not be directly compared (detector analysis may still be applied, but attention must be given to changes in sensitivities).

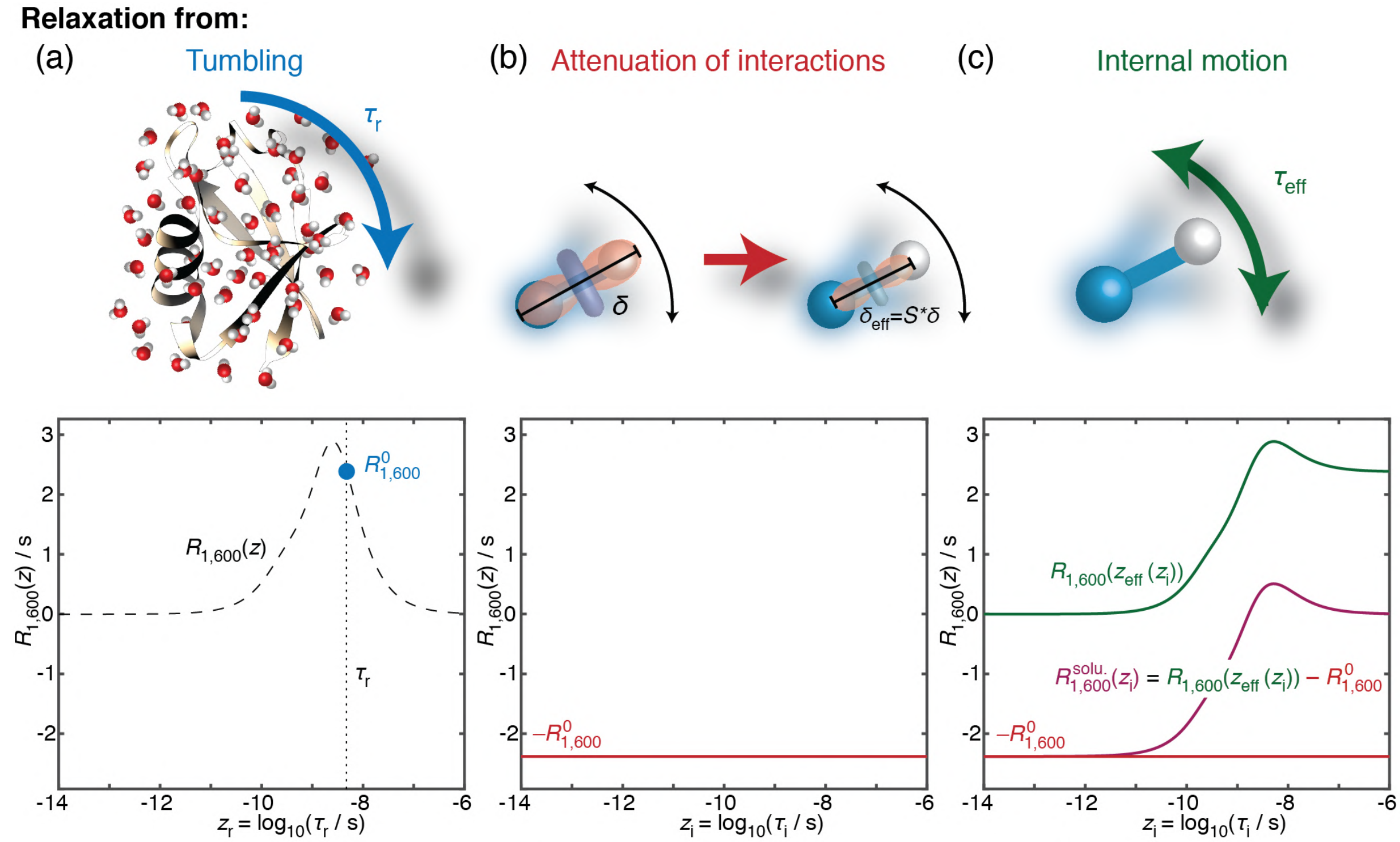

$$\text{(d)}\quad R^{(\theta,S)}_{1,600} = R^0_{1,600} + (1-S^2)\int_{-\infty}^{\infty}\theta(z_i)\underbrace{(R_{1,600}(z_{eff}(z_i)) - R^0_{1,600})}_{R^{solu.}_{1,600}(z_i)}dz_i$$

**Fig. 4.** Contributions to the $^{15}$N $R_1$ relaxation-rate constant at 600 MHz with $\tau_r$ = 4.84 ns. (a) Relaxation due to tumbling for an internally rigid molecule may be calculated by evaluating $R_{1,600}(z_r) = R^0_{1,600}$, where $z_r = \log_{10}(\tau_r / 1\,\mathrm{s})$. Then, (a) shows $R_{1,600}(z)$ as a dashed line, and $z_r = \log_{10}(\tau_r)$ as vertical, dotted line, with the resulting $R^0_{1,600}$ shown as a blue circle. $R^0_{1,600}$ appears as a constant offset for calculation of the relaxation-rate constant, $R^{(\theta,S)}_{1,600}$ (see (d)). (b) Internal motion results in a reduction of the effective size of anisotropic interactions, such that $\delta_{eff} = S\delta$ ((b), top), yielding a reduction in relaxation by $(1-S^2)R^0_{1,600}$. This reduction is scaled by the total internal motion, $(1–S^2)$, but does not depend on the correlation time, resulting in a uniform, negative contribution to the sensitivity to internal motion of $-R^0_{1,600}$. (c) The effective internal motion (internal motion composed with tumbling) induces some relaxation directly, although with an effective correlation time ( $z_{eff} = \log_{10}(\tau_{eff} / 1\,\mathrm{s})$ ), illustrated in (c) with $R_{1,600}(z_{eff}(z_i))$ plotted ($z_{eff}$ is a function of $z$ and $z_r$, see Eq. (16) and Fig. 3). The sensitivity to internal motion, $R^{solu.}_{1,600}(z_i)$, is finally obtained by summing $R_{1,600}(z_{eff}(z_i))$ and $-R^0_{1,600}$, which is plotted in magenta. This function along with the distribution of motion ($(1-S^2)\theta(z_i)$) may then be used to calculate the relaxation rate constant, $R^{(\theta,S)}_{1,600}$, as given in (d). Note that for correlation times much longer than the tumbling correlation time, the terms $-R^0_{1,600}$ and $R_{1,600}(z_{eff}(z_i))$ cancel out, illustrating the fact that the tumbling masks the influence of motions with correlation times much longer than the correlation time of the tumbling.

We can decompose the contributions to the relaxation-rate constant given in Eqs. (18) and (19) into three parts, as illustrated in Fig. 4, which depend on the internal distribution of motion, $(1–S^2)\theta(z_i)$, and the correlation time of the tumbling, $\tau_r$: (i) Relaxation induced by tumbling alone, as in the case of a completely rigid molecule. (ii) Reduction of relaxation from tumbling, due to attenuation of NMR interactions by internal motion. (iii) Relaxation induced directly by the effective internal motion (see Eq. (18)).

The separation into three contributions seems at first slightly counterintuitive: we expect tumbling to mask the influence of motions with correlation times significantly longer than $\tau_\mathrm{r}$. The attenuation of NMR interactions by internal motions ($\delta_\mathrm{eff} = S\delta$) can be considered uniform for all correlation times (subtracting $R_\zeta^0$ from the sensitivity), while relaxation induced directly by internal motion (adding $R_\zeta(z_{eff}(z_i))$ to the sensitivity) depends on $z_\mathrm{i}$ but approaches $R_\zeta^0$ for long correlation times. However, the sum of the latter two contributions is 0 for long correlation times, yielding the expected behavior. This is equivalent to the usual description: internal motions much slower than the overall tumbling are not relaxation active.

In principle, it is also possible to characterize the solution-state relaxation-rate constants using the methodology developed for solid-state NMR. However, such an analysis would provide the total distribution of motion, $\theta_\mathrm{tot.}(z)$, which yields the correlation function via Eq. (8). This distribution describes the internal motion (having an effective correlation time as opposed to the internal correlation time) and the overall tumbling motion. Since the overall tumbling leads to an isotropic distribution of orientations, the order parameter is then $S^2 = 0$, such that $(1–S^2) = 1$, and we obtain

$$\begin{aligned} R_\zeta^{(\theta,S)} &= R_\zeta^0 + (1-S^2)\int_{-\infty}^{\infty}\theta(z_\mathrm{i})R_\zeta^\mathrm{solu.}(z_\mathrm{i})dz_\mathrm{i} \\ &= R_\zeta^{(\theta_\mathrm{tot.},0)} = \int_{-\infty}^{\infty}\theta_\mathrm{tot.}(z)R_\zeta(z)dz \end{aligned} \quad . \qquad (20)$$

The distribution of total motion, $\theta_\mathrm{tot.}(z)$, is different from the distribution of internal motion, $(1-S^2)\theta(z_\mathrm{i})$, since overall tumbling ($z$ = $z_\mathrm{r}$) and internal motion ($z$ = $z_\mathrm{eff}$) contribute to the distribution of total motion (for $\theta_\mathrm{tot.}(z)$, $z$ can be both the tumbling correlation time, $z_\mathrm{r}$, or $z_\mathrm{eff}$, resulting from internal motion and tumbling). Note that there is a well-defined relationship between the two distributions, given in the SI Section 1.

We investigate the behavior of the sensitivity to internal motion ($R_\zeta^\mathrm{solu.}(z_\mathrm{i})$) by considering several typical sets of experiments. For example, Fig. 5(a) shows the normalized sensitivities, $R_\zeta(z)$, to the distribution of the total motion, $\theta_\mathrm{tot.}(z)$, for $R_1$, $R_2$, and $\sigma_\mathrm{NH}$ rate constants. In Fig.

5(b), normalized sensitivities, $R_\zeta^{\text{solu.}}(z_\text{i})$, to the distribution of the internal motion, $(1-S^2)\theta(z_\text{i})$, are given. We see a number of differences: first at short correlation times, the sensitivities to internal motion, $R_\zeta^{\text{solu.}}(z_\text{i})$, are negative due to the correction term $R_\zeta^0$ (see Eq. (18)). At sufficiently short correlation times, very little relaxation is induced directly by internal motion, and the sensitivity is dominated by the term $-R_\zeta^0$ in Eq. (18), resulting in a reduction of the relaxation-rate constant compared to a rigid molecule. At longer correlation times, the sensitivity to internal motion increases and, in some cases, becomes positive, but when the internal correlation time becomes larger than the correlation time of the overall tumbling, all the $R_\zeta^{\text{solu.}}(z_\text{i})$ approach zero, since the tumbling masks internal motions that are significantly slower than the tumbling. The sensitivity of the $R_1$ rate constant varies significantly in the range 600 to 950 MHz (Fig. 5(c)), with less variation for the sensitivity $\sigma_{\text{NH}}$, and almost no variation for $R_2$ in the same range of magnetic fields.

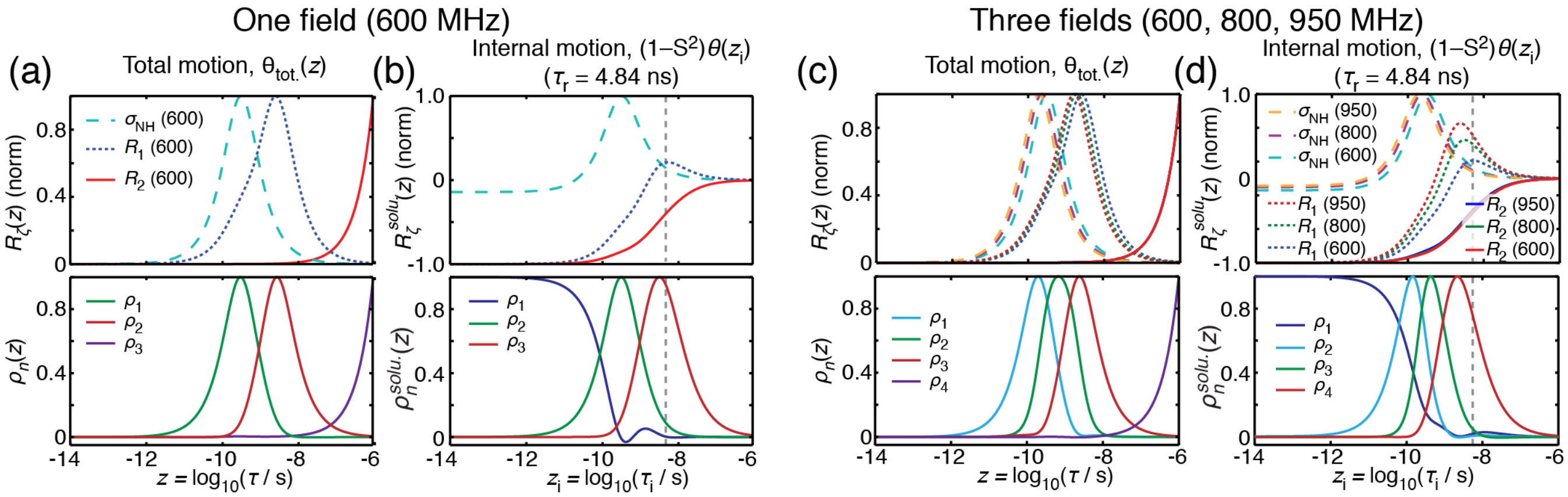

**Fig. 5.** Experimental sensitivities for total ($R_\zeta(z)$) and internal ($R_\zeta^{\text{solu.}}(z_\text{i})$) motions, and optimized detector sensitivities ($\rho_n(z)$). (a) Experimental sensitivities, $R_\zeta(z)$, for $\sigma_{\text{NH}}$, $R_1$, and $R_2$ rate constants at 600 MHz for the total motion (top, with sensitivities normalized to 1), and an optimized set of detector sensitivities, $\rho_n(z)$, obtained by linear combination of the rate constant sensitivities (bottom). (b) Sensitivity to internal motion, $R_\zeta^{\text{solu.}}(z_\text{i})$, calculated assuming a tumbling correlation time of $\tau_r$ = 4.84 ns (top) and resulting detector sensitivities, $\rho_n^{\text{solu.}}(z_\text{i})$ (bottom). (c) shows experimental sensitivities, $R_\zeta(z)$, at 600, 800 MHz, and 950 MHz for the total motion (top), with a set of four detector sensitivities, $\rho_n(z)$, calculated (bottom). (d) The same set of experiments with sensitivities to internal motion $R_\zeta^{\text{solu.}}(z_\text{i})$ ($\tau_r$ = 4.84 ns). In (b) and (d), $\tau_r$ is indicated with a grey dotted line through all plots. In each section, the experimental sensitivities are normalized, so that the maximum of the absolute value is 1. Note that the normalization of the $R_2$ sensitivity to the total motion, $R_2(z)$, is determined by the longest correlation time in the plot, so that decreasing the maximum $z$ in plots in (a) and (c) would cause $R_2(z)$ to appear to shift to the left (correspondingly, $\rho_3(z)$ would also shift).

### C. Optimized linear combinations for detector design

Dynamics detectors are generated by optimizing a linear combination of the relaxation-rate constant sensitivities to obtain detector sensitivities, which are well separated into different ranges of correlation times. Optimized linear combinations for the relaxation-rate constant sensitivities shown in Fig. 5(top) were generated and plotted in Fig. 5(bottom). Note that optimization methods discussed for solid-state NMR[13] are applicable to those used in solution-state NMR, despite the appearance of negative sensitivities.

The detector sensitivities of total and internal motion differ markedly for short and long correlation times (Fig. 5(a)/(c) vs. (b)/(d), bottom). Considering the analysis of relaxation-rate constants measured at a single magnetic field (Fig. 5(a) and (b)) we find two of the three detectors in approximately the same positions for total and internal motion (corresponding detectors are plotted with the same color). However, the third detector ($\rho_3$ for total motion, $\rho_1$ for internal motion) has moved significantly. $\rho_3(z)$ for the total motion is strongly dependent on $R_2$, and diverges as one approaches long correlation times, whereas $\rho_1^{\text{solu.}}(z_\text{i})$ of the internal motion is nearly uniformly sensitive at short correlation times. Differences arise because the detectors characterize different distributions of motion (internal motion, $(1–S^2)\theta(z_\text{i})$, or total motion $\theta_{\text{tot.}}(z)$, see SI section 1 for comparison). The sensitivity to internal motion is altered by the tumbling, which masks slow motions. As a consequence, the sensitivities, $\rho_n^{solu.}(z_\text{i})$ must approach zero above the overall tumbling correlation time. On the other hand, $\rho_1^{\text{solu.}}(z_\text{i})$ is sensitive to motion at short correlation times, since one can determine how much the measured relaxation-rate constants have been reduced from the expected relaxation for an internally rigid molecule (due to attenuation of the effective size of anisotropic interactions). Note that this can be determined because we consider the correlation time of the tumbling determined independently. The overall tumbling correlation time must be determined before detector analysis, using existing methods, e.g. the program ROTDIF [29] was used in this study.

Similar behavior is observed in the analysis of relaxation at three magnetic fields (Fig. 5(c)/(d), bottom). In principle, up to nine detectors can be optimized for nine relaxation rate constants. However, $R_2$ sensitivities are typically very similar so that one rarely gains additional discrimination between correlation-time ranges by using more than one $R_2$ experiment. Multiple $R_2$ experiments are nonetheless useful because they increase signal to noise and allow the determination of contributions of broadening due to fast chemical exchange to $R_2$ (see below). Similarly, multiple high-field $R_1$ usually only provide two detectors, as do multiple NOE experiments (although sufficient separation in $B_0$ fields and signal-to-noise may allow more). Such a three-field data set can be used to optimize three to five detectors, depending on the signal-to-noise ratio and the separation of the $B_0$ fields. Here, we have optimized four detectors (SI section 2.3 discusses the choice of number of detectors). The range of sensitivities barely increases from one to three fields (considering detectors $\rho_2$–$\rho_4$; $\rho_1$ is always sensitive to the shortest correlation times for solution-state data). This is because the shortest correlation times to which $\rho_2$ is sensitive is determined by the highest field at which the NOE ($\sigma_{NH}$) was measured, and the sensitivity to long correlation times is limited by the rotational correlation time (as opposed to the choice of the experimental parameters). So, it is possible to shift $\rho_2^{\text{solu.}}(z_i)$ towards shorter correlation times, by using a larger $B_0$ field for the NOE experiment, but sensitivity to longer correlation times can only be significantly increased if the rotational correlation time becomes longer and the magnetic field, $B_0$, lower. Changing the $B_0$ field within the range of high fields used in biomolecular NMR has very limited effect, since the variations of sensitivity of different $\sigma_{NH}$ is relatively small compared to the difference in sensitivity of $\sigma_{NH}$ and $R_1$ at the same field.

We have previously developed a graphical method of optimization to generate detector sensitivities from linear combinations of the relaxation-rate constant sensitivities, using "allowed spaces".[13] Here we simply review the definition of the spaces and how they are used to generate the linear combination of relaxation-rate constant sensitivities. An allowed space can be understood as follows: suppose we record a set of $N$ experiments. Then, we can take an $N$-dimensional space, where each axis represents the value of one of the relaxation-rate constants. Not all combinations of relaxation-rate constants are physically possible given an arbitrary distribution of motion, $(1-S^2)\theta(z)$, so that we may determine what points in the space correspond to a set of relaxation-rate constants that can result from some distribution of motion. All possible sets of rate constants for an arbitrary distribution of motion are then referred to as the "allowed space". Note that for solution-state relaxation, a molecule with no internal motion will still have non-zero relaxation-rate constants, due to overall tumbling (Eq. (17)). Therefore, when plotting the allowed space for solution-state relaxation, we first calculate $R_\zeta^{(\theta,S)} - R_\zeta^0$, so that the origin of the space corresponds to no internal motion ($1-S^2 = 0$). We also use rate constants with normalized axes denoted as $\mathfrak{R}_\zeta^{(\theta,S)}$, where $\zeta$ indicates the experiment, to yield

$$\begin{aligned}&\Re_\zeta^{(\theta,S)} = (R_\zeta^{(\theta,S)} - R_\zeta^0)/c_\zeta \\ &c_\zeta = \text{median}(\sigma(R_\zeta)) \\ &\quad \textit{or} \\ &c_\zeta = \max\left|R_\zeta^{\text{solu.}}(z)\right|\end{aligned} \qquad (21)$$

In the case that one plots the allowed space for a particular set of experimental measurements, $c_\zeta$ can be taken to be the standard deviation of the measurement of rate constant, $R_\zeta^{(\theta,S)}$, or its median for a rate constant measured at multiple sites. The distance between two points in the allowed space quantifies how easily these points may be distinguished from the given experimental data set. In the absence of experimental data, one can take the maximum of the absolute value of the sensitivity to internal motion, so that all experiments are on a similar scale.

The allowed space of relaxation-rate constants for a data set including $R_1$, $R_2$, and $\sigma_{IS}$ at a single field (at 600 MHz, taking $c_\zeta = \max\left|R_\zeta(z)\right|$), was computed (see Fig. 6). The origin corresponds to no internal motion ($1-S^2 = 0$). The observed relaxation-rate constants at the origin are non-zero, due to the offset terms, $R_\zeta^0$, as indicated in Eq. (21). Positions in the space that can result from internal motion with a single correlation time (Dirac distribution) are shown as solid lines (see Fig. 6 with $1-S^2 = 1$ and $1-S^2 = 0.5$). The volume shown corresponds to any point that can be constructed from a (positive) linear combination of positions in the space corresponding to single correlation times, i.e. any point that can result from some distribution of internal motion, $(1-S^2)\theta(z_\mathrm{i})$.

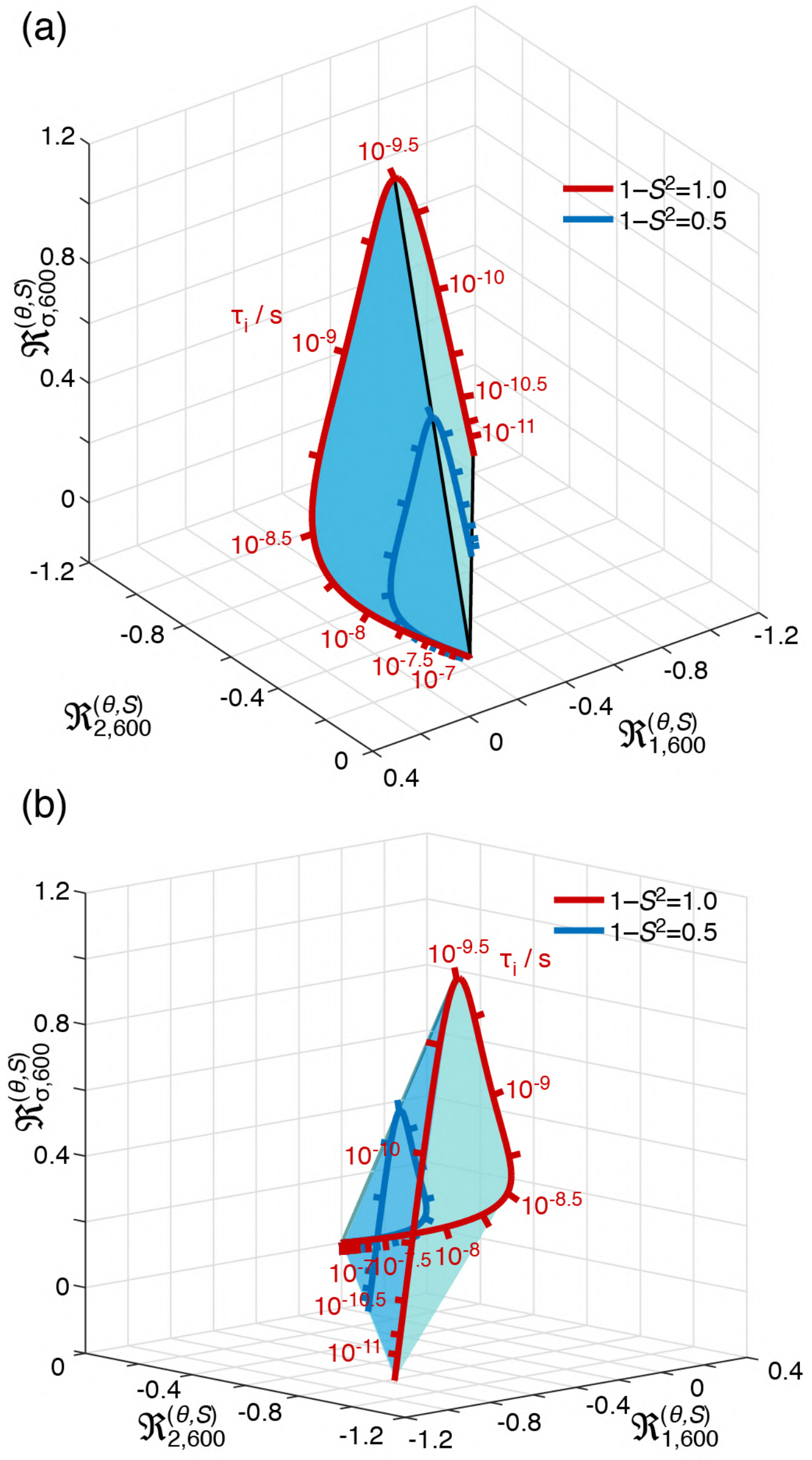

**Fig. 6.** Allowed space of normalized rate constants for $^{15}$N $R_1$, $R_2$, and $\sigma_{HN}$ rate constants acquired at 600 MHz, assuming $\tau_r$ = 4.84 ns. Two views are shown in (a) and (b), where the axes are the normalized rate constants, $\mathfrak{R}_{\zeta}^{(\theta,S)}$. Sets of the three rate constants which are possible for an arbitrary distribution of internal motion, (1–$S^2$)$\theta(z_i)$ are highlighted in blue (allowed space, different shading shows different sides of the space). Traces show positions in the space corresponding to exactly one correlation time, with the red trace having an order parameter, $S^2$, such that (1–$S^2$) = 1, and blue having an order parameter such that (1–$S^2$) = 0.5. Note that the allowed space is a volume and contains all points that are along the red trace, and additionally all points that are between two or more points on the red trace.

We note that for a given data set, the information about how motion is distributed over different internal correlation times, as described by $\theta(z_i)$, is contained entirely in the ratios of the various rate constants, whereas the total amplitude of motion, (1–$S^2$), is obtained from the magnitude of the rate constants. Therefore, we have introduced a “reduced space” of rate constants, for which we define a ratio of the relaxation-rate constants, in order to remove dependence on the total amplitude of motion (reducing the dimensionality has practical advantages, in particular allowing one to visualize the allowed space of rate constants for three rate constants in a 2D plot). Previously, we have defined the dimensions of the reduced space to be given by some $\kappa_\zeta = \Re_\zeta^{(\theta,S)} / \Sigma_\zeta \Re_\zeta^{(\theta,S)}$, where $\Sigma_\zeta \Re_\zeta^{(\theta,S)}$ indicates the sum of all normalized relaxation rate constants. For $N$ experiments, one obtains then $N$–1 linearly independent $\kappa_\zeta$ to define the reduced space. When defining the $\kappa_\zeta$ for solution-state analysis, however, we must be careful because the $\Re_\zeta^{(\theta,S)}$ can be both negative and positive, so that $\Sigma_\zeta \Re_\zeta^{(\theta,S)}$ may cross zero, causing $\kappa_\zeta$ to diverge at such points. Therefore, we use one of the experiments, $\zeta$, for which the corresponding sensitivity, $R_\zeta^{\text{solu.}}(z_\text{i})$, remains negative at all values of $z_\text{i}$ to define the reduced space. Such a behavior is often observed for relaxation-rate constants which sample the spectral density at zero frequency ($J(0)$), i.e., transverse relaxation-rate constants. For the example shown above with relaxation-rate constants $R_1$, $R_2$, and $\sigma_{\text{NH}}$ at 600 MHz, the corresponding reduced space can be defined by dividing by $\Re_{2,600}^{(\theta,S)}$ so that

$$\kappa_{R1,600} = \frac{\Re_{R1,600}^{(\theta,S)}}{-\Re_{R2,600}^{(\theta,S)}}, \quad \kappa_{\sigma,600} = \frac{\Re_{\sigma,600}^{(\theta,S)}}{-\Re_{R2,600}^{(\theta,S)}}, \tag{22}$$

where the dimensionality of the reduced space is one less than the number of experiments. An example of the reduced space is shown for $R_1$, $R_2$, and $\sigma_{\text{NH}}$ at 600 MHz, for both the total motion (includes tumbling in solution) and the internal motion (tumbling removed) in Fig. 7(a) and (b), respectively.

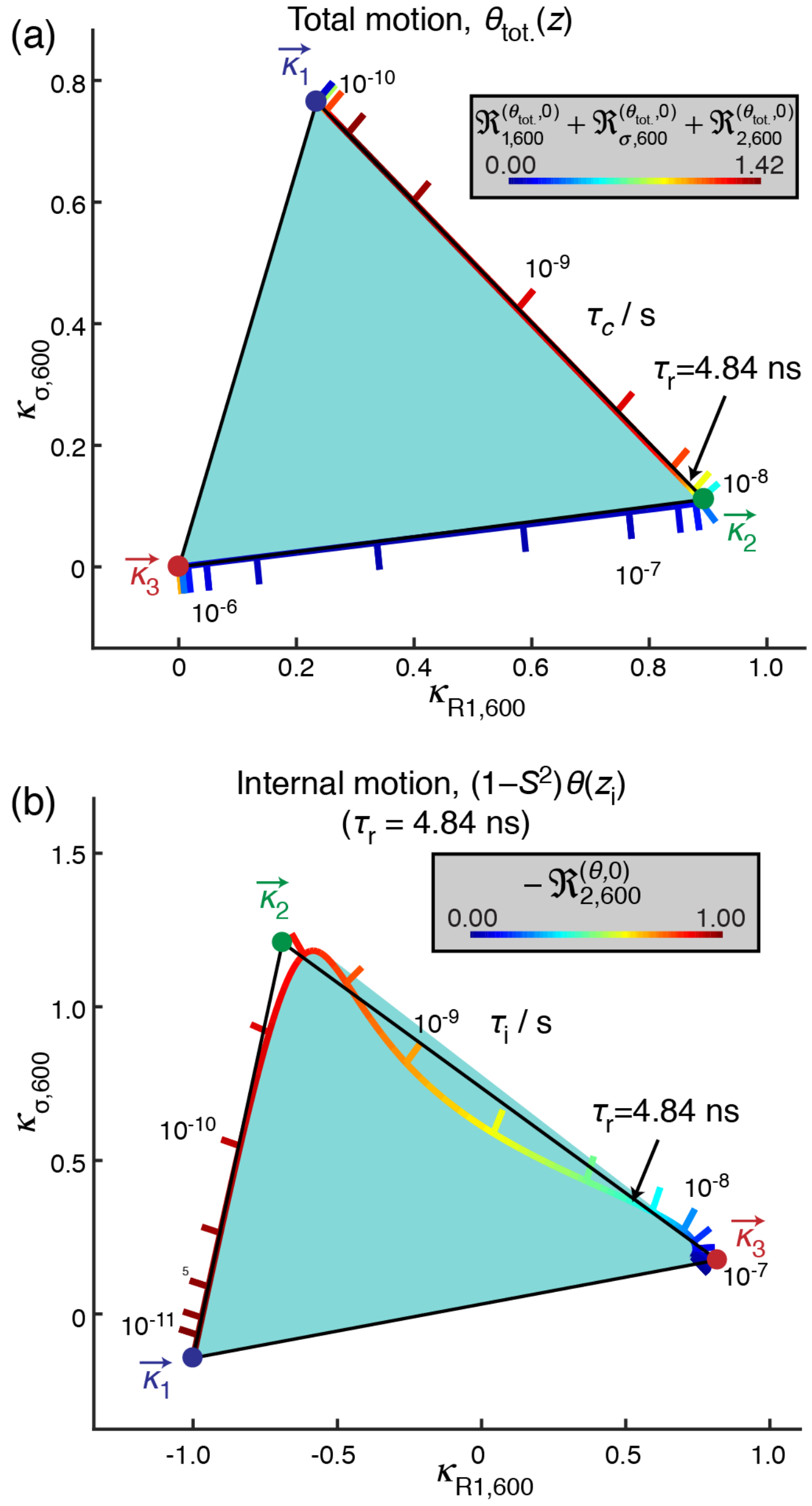

**Fig. 7.** Reduced space of normalized rate constants for $R_1$, $R_2$, and $\sigma$ rate constants at 600 MHz, where the *x*-, and *y*-axes correspond to $\kappa$. (a) Allowed region (cyan) for the sensitivities to the total motion (for characterizing $\theta_{\mathrm{tot.}}(z)$, see Eq. (20)), where the $\kappa$ are obtained by dividing by $\Re_{1,600}^{(\theta_{\mathrm{tot.}},S)}+\Re_{\sigma,600}^{(\theta_{\mathrm{tot.}},S)}+\Re_{2,600}^{(\theta_{\mathrm{tot.}},S)}$ (this value is color-coded onto the plot for $S$=0 when the position corresponds to a single correlation time). (b) Allowed region (cyan) for the sensitivities to the internal motion ($(1\text{–}S^2)\theta(z_i)$, assuming $\tau_r$ = 4.84 ns), where the $\kappa$ coefficients are obtained by dividing by $-\Re_{2,600}^{(\theta,S)}$ (value color-coded onto the plot for $S$=0 when the position corresponds to a single internal correlation time). In both (a) and (b), good positions for the $\vec{\kappa}_n$ are shown as colored dots, which indicate the direction of the detection vectors ($\vec{r}_n$). These correspond to the sensitivities shown in Fig. 5(a) and (b), respectively, after applying normalization (for example, see Eq. (23)).

Detectors are generated by selecting an optimal set of "detection vectors" that extend into the full space. Correspondingly, these are points in the reduced space (their positions denoted as $\vec{\kappa}_n$). From these positions, it is possible to determine the direction of the detection vector, in this example defined by $\kappa_{R1,600}$ and $\kappa_{\sigma,600}$, according to

$$\vec{r}_n = a_n \begin{pmatrix} \kappa_{R2,600} c_{R2,600} \\ \kappa_{R1,600} c_{R1,600} \\ \kappa_{\sigma,600} c_{\sigma,600} \end{pmatrix}. \qquad (23)$$

$$\kappa_{R2,600} = -1$$

Recall that the $\kappa_\zeta$ define ratios of the rate constants, but not their absolute values, so that a point in the reduced space ($\vec{\kappa}_n$) does not define the length of the detection vector, only its direction. The length is then determined by adjustment of $a_n$, which changes the amplitude of the corresponding detector sensitivity since it is inversely proportional to $a_n$ (as discussed previously;[13] we use the equal-maximum normalization here, with all sensitivities having maxima of one). Ideally, one surrounds (or nearly surrounds) the reduced space with a minimal number of $\vec{\kappa}_n$. To fully surround the space, it is necessary to have at least $N$ different $\vec{\kappa}_n$ for $N$ experiments. However, one may also reduce the number of $\vec{\kappa}_n$, yielding fewer detectors, but obtain a more precise determination of the remaining detectors.[13] The colored dots in Fig. 7(a) and (b) indicate good choices for $\vec{\kappa}_n$ to yield well-separated detector sensitivities, for the total motion (solid-state) and internal motion (solution-state), respectively. The positions yield the detector sensitivities shown in Fig. 5(a) and (b) (bottom).

As in Eq. (12), measured relaxation-rate constants in solution-state are fitted to detection vectors, $\vec{r}_n$. For solution-state data, due to the offset term, $R_\zeta^0$, appearing in Eq. (21), the calculated detector responses are given by

$$\begin{pmatrix} \rho_1^{(\theta,S)} \\ \vdots \\ \rho_n^{(\theta,S)} \end{pmatrix} = \begin{pmatrix} [\vec{r}_1]_\zeta / \sigma(R_\zeta) & \cdots & [\vec{r}_n]_\zeta / \sigma(R_\zeta) \\ \vdots & \ddots & \vdots \\ [\vec{r}_1]_\xi / \sigma(R_\xi) & \cdots & [\vec{r}_n]_\xi / \sigma(R_\xi) \end{pmatrix}^{-1} \begin{pmatrix} (R_\zeta^{(\theta,S)} - R_\zeta^0) / \sigma(R_\zeta) \\ \vdots \\ (R_\xi^{(\theta,S)} - R_\xi^0) / \sigma(R_\xi) \end{pmatrix}. \qquad (24)$$

Here, the variables $\zeta$ to $\xi$ span the experimental data set (e.g. for a one field data set at 600 MHz, the $\zeta$, $\xi$ would be replaced by $R_{2,600}$, $R_{1,600}$, $\sigma_{600}$). Before fitting one subtracts $R_\zeta^0$ from each experimental rate constant. Note that the number of detection vectors cannot exceed the number of experiments, and in practice there are usually fewer detection vectors than experiments. In particular when experiments have similar sensitivities, the use of too many detection vectors would result in some of them being almost co-linear, so that the matrix shown in Eq. 24 would be almost singular (i.e., lacking an inverse) increasing the error of the analysis (see SI section 2.2 for more details). One also obtains the detector sensitivities from the detection vectors, which results in a similar expression as in Eq. (24).

$$\begin{pmatrix} \rho_1^{solu.}(z) \\ \vdots \\ \rho_n^{solu.}(z) \end{pmatrix} = \begin{pmatrix} [\vec{r}_1]_\zeta / \sigma(R_\zeta) & \cdots & [\vec{r}_n]_\zeta / \sigma(R_\zeta) \\ \vdots & \ddots & \vdots \\ [\vec{r}_1]_\xi / \sigma(R_\xi) & \cdots & [\vec{r}_n]_\xi / \sigma(R_\xi) \end{pmatrix}^{-1} \begin{pmatrix} R_\zeta^{solu.}(z)/\sigma(R_\zeta) \\ \vdots \\ R_\xi^{solu.}(z)/\sigma(R_\xi) \end{pmatrix} \tag{25}$$

Note that we have modified Eq. (25) slightly from its previous form, where normalization by the standard deviations, $\sigma(R_\zeta)$ was not indicated.[13] This usually makes little difference in the resulting sensitivities, but is a more rigorous definition in the case that standard deviations are included when fitting the rate constants as in Eq. (24).

Although allowed spaces may be used for visualization of the information content of a relaxation data set, and subsequent placement of detection vectors, $\vec{r}_n$ (via the placement of $\vec{\kappa}_n$ in the reduced space) to generate optimized linear combinations of rate constants, this method may become cumbersome for large data sets. A solution is to use singular value decomposition[30] for detector optimization. (see Supplementary information section 2.1). One can also estimate detector uncertainties as a function of the resulting singular values (Supplementary Information, Sections 2.2, 2.3). Tools to perform this optimization and subsequent analysis are provided in DIFRATE version 2,[24] which is available for MATLAB (also available without a MATLAB license via MATLAB Runtime). In the analysis of typical $^{15}N$ relaxation data sets presented below, we have used this improved approach.

## D. Correcting for exchange contributions

Thus far, we have assumed that all contributions to the measured relaxation-rate constants can be explained by the distribution of motion (Eq. (20)), describing internal stochastic motion, and by overall tumbling of the molecule in solution. However, other sources of relaxation may exist, in particular the contribution of exchange to transverse relaxation rates, $R_2$. In this case, we must also account for such a process in our analysis. The analysis of ps-ns motion can be performed with data at multiple magnetic fields. If the exchange process is in the fast-exchange regime ($2\pi(\nu_1-\nu_2)\tau_{ex} << 1$), where $\nu_1$ and $\nu_2$ are the two resonance frequencies of the exchanging resonance, $R_2$ is proportional to $(\nu_1-\nu_2)^2$, which is in turn proportional to $B_0^2$. In this case, one can add an additional detection vector with non-zero terms corresponding to each $R_2$ experiment, which are proportional to $B_0^2$. For example,

$$\vec{r}_{ex} = \begin{pmatrix} 0 \\ \vdots \\ B_{0,\xi}^2 \\ \vdots \\ B_{0,\zeta}^2 \\ 0 \\ \vdots \end{pmatrix} / B_{0,\xi}^2 \tag{26}$$

could be added to a set of detection vectors where the $B_{0,\zeta}^2$ give the static magnetic fields of the $R_2$ experiments. Then, this detection vector is also fitted to the data with the rest of the detection vectors, and will fit deviations of $R_2$ relaxation behavior due to fast exchange contributions. Normalization of this detection vector will not affect our ability to factor out the influence from chemical exchange. However, in the normalization scheme here, we set one of the elements to one, so that the responses of this detector will estimate $R_{2,\mathrm{ex}}$ at the field corresponding to this element. This method of accounting for chemical exchange is only applicable with $R_2$ acquired at multiple fields. Note that there is no corresponding sensitivity function ($\rho_n(z)$) for this detector.

## III. Results and Discussion

We have applied detectors derived from simple one-field or typical multi-field data sets (three fields) to relaxation data previously acquired on ubiquitin in solution-state NMR,[31] (Fig. 8). The rotational correlation time was determined previously, using the ROTDIF software.[29] The analysis of relaxation data acquired at two fields is shown in Supplementary information Section 3. The results obtained with relaxation rates measured at one or three magnetic fields are similar. $\rho_1$ (<~100 ps) yields relatively uniform behavior for both one- and three-field data sets, with more motion at the C-terminus. $\rho_3$ (one field, ~4 ns) and $\rho_4$ (three fields, ~3 ns) also exhibit similar behavior for both analyses (we indicate the approximate center of the detector in parentheses, where the widths cover just over an order of magnitude). Uncertainties are slightly smaller for $\rho_1$ and significantly smaller for $\rho_4$ in the three-field analysis ($\rho_4$ compared to $\rho_3$ in the one-field analysis), which simply results from the use of more data (and therefore better signal-to-noise) in the three-field analysis and not the inclusion of new information. $\rho_2$ (one-field, ~250 ps) and $\rho_2$/$\rho_3$ (three-field, ~100/500 ps) show increased motion around residues 7-13 (β1-β2 turn), as well as more motion at the C-terminus, with relatively little motion elsewhere. Motion measured with $\rho_2$ when using only one field is split between the two detectors $\rho_2$ and $\rho_3$ when combining data from three fields, although splitting this detector results in larger uncertainties (the choice of number of detectors using three fields is investigated with variants of the Akaike information criterion[32-36] in SI section 4). In the multi-field data set, we have also accounted for exchange contributions to $R_2$ relaxation,[1] by including an additional detector that fits fast exchange (such that $R_{2,ex} \propto B_0^2$). This removes several distortions due to exchange, appearing primarily in $\rho_1^{(\theta,S)}$ (res. 23, 25, 70). Overall, we obtain an accurate dynamics detector analysis with separation of ranges of correlation times from typical high-field data sets.

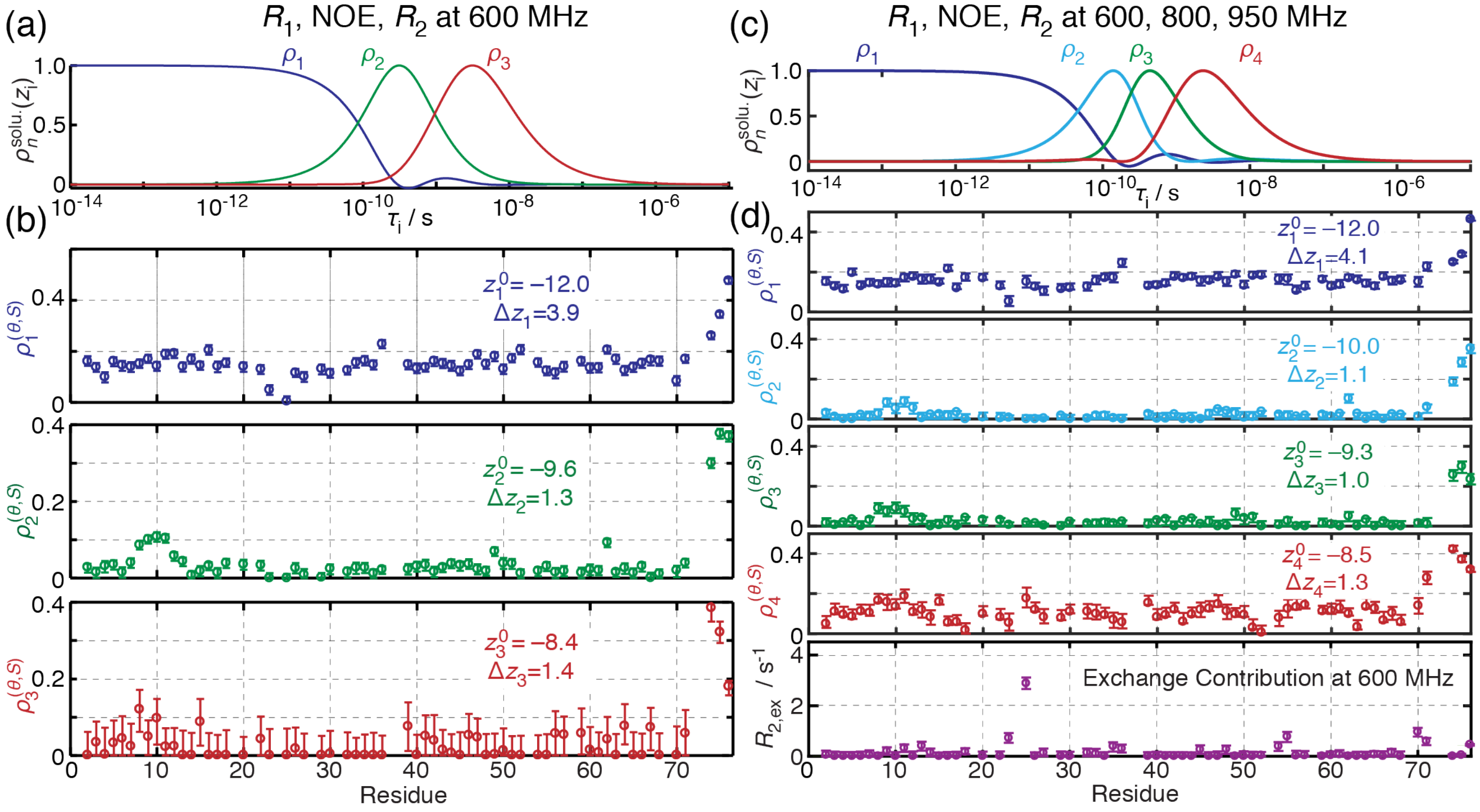

**Fig. 8.** Detector responses for ubiquitin from $R_1$, $\sigma_{NH}$, and $R_2$ relaxation-rate constants acquired at one or three magnetic fields. (a) shows the detector sensitivities ($\rho_n^{solu.}(z_i)$) calculated from $R_1$, $\sigma_{NH}$, and $R_2$ rate constants at one field (600 MHz, definition of detection vectors in SI Table S2). (b) shows the experimental detector responses from data at this single field. (c) shows sensitivities calculated from relaxation-rate constants measured at three fields (600, 800, 950 MHz, definition of detection vectors in SI Table S4). (d) shows the detector responses from relaxation data measured at these three fields, and also shows the fitted exchange contribution (plotted value corresponds to 600 MHz, where $R_{2,ex} \propto B_0^2$). Error bars indicate the 95% confidence interval, determined by Monte Carlo error analysis (200 repetitions).[13] Each plot in (b) and (d) indicates $z_0$ and $\Delta z$, which are the center of the detector and the effective width of the detector, which approximate the average correlation time and the range of correlation times a detector is sensitive to (both on log-scales, with precise definitions given in the SI glossary). Data fits are found in Supplementary information Figs. S7 and S9.

In the current analysis, we have neglected the anisotropy of the rotational diffusion tensor. Previously, the anisotropy under these experimental conditions was determined to be small with $D_{\parallel} / D_{\perp} = 1.18$,[31] so that the overall correlation function ($C_{\mathrm{O}}(t)$) decays slightly slower for H–N bonds in Ubiquitin parallel to the *z*-component of the diffusion tensor in its principle axis system (PAS), and slightly faster for H–N bonds perpendicular to the *z*-component. This means that the correction terms, $R_{\zeta}^{0}$, may not fully remove all relaxation contributions due to overall tumbling or may remove too much, depending on bond orientation. This difference in the relaxation could be wrongly interpreted as internal dynamics. A treatment that substitutes the overall correlation function in Eq. (13) with a correlation function for anisotropic tumbling would improve the analysis, especially for molecules with larger anisotropies of the rotational diffusion tensor. We are currently implementing such a scheme, which is beyond the scope of the present paper.

We have investigated whether our results were significantly biased by this simplification. The amount of relaxation due to tumbling depends directly on the orientation of the individual H–N bond vectors relative to the diffusion tensor (strictly speaking the orientation of the H–N dipole coupling and $^{15}$N CSA tensors, which are not exactly aligned). For example, $R_2$ relaxation due to tumbling should be faster where the bond vectors point along the *z*-axis, since rotational diffusion around the *z*-axis is slower. Therefore, we plot the square of the *z*-component of the bond vector in the PAS of the diffusion tensor (the bond vector is normalized to a length of 1, and the square is relevant for relaxation). Results are shown in Fig. 9; there appears to be some correlation, so that when $[v_{\mathrm{HN}}]_z^2$ becomes small, $\rho_4^{(\theta,S)}$ increases (res. 18-20, 35-36, especially 51-54, 63). This is somewhat expected as, for H–N bond vectors perpendicular to the z-axis of the PAS, tumbling motion is slightly faster than the overall correlation time, inducing faster $R_1$ relaxation; this additional relaxation is underestimated in the correction by the term $R_{\zeta}^{0}$, and then the increased relaxation rate increases $\rho_4^{(\theta,S)}$. Although the effect is weak, it will be necessary to improve the diffusional model for systems with larger anisotropies to avoid significant distortions.

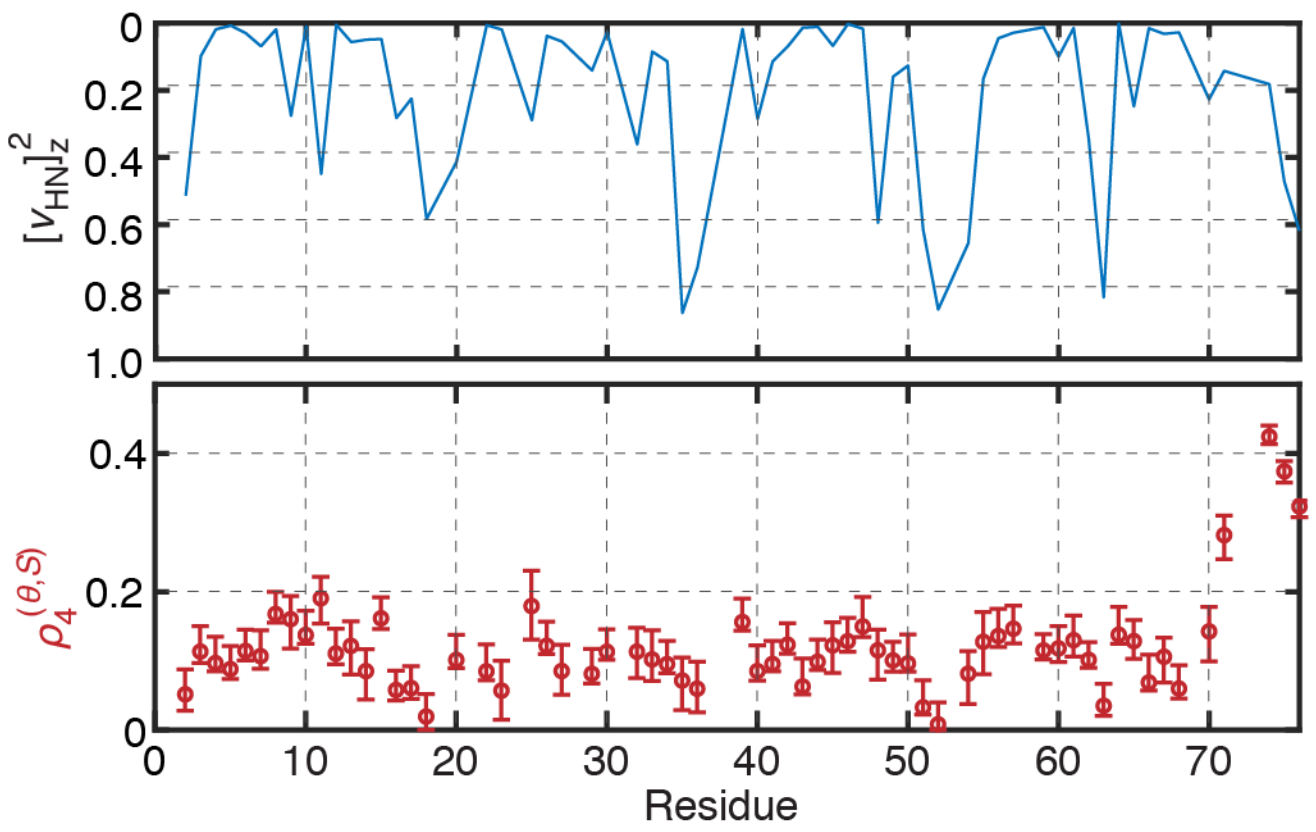

**Fig. 9.** Square of the component of the H–N bond vectors parallel with the *z*-component of the diffusion tensor. There is weak correlation between increases in $[v_{\mathrm{HN}}]_z^2$ and decreases in $\rho_4^{(\theta,S)}$, as seen in the comparison here ( $[v_{\mathrm{HN}}]_z^2$ plotted with an inverted axis for better comparison. $\rho_4^{(\theta,S)}$ values are the same as those shown in Fig. 8(d).

## A. Comparison to model-free analysis

The detector analyses (Fig. 8) may be compared to a model-free/extended model-free analysis of relaxation datasets recorded at three magnetic fields, which is shown in Fig. 10. The model-free analysis was performed by Charlier et al.,[31] using the program DYNAMICS.[37] Model-free analysis displays some discontinuity of the fitted parameters along the primary sequence. Discontinuity appears to have two primary sources. The first is model selection: relaxation data is analyzed in this example with four different models of the correlation function. One uses either one or two motions in the model, and in some cases it is assumed that the correlation time of the faster motion is too short to directly induce relaxation, so that only its amplitude is fitted. Then, models with anywhere from one to four parameters are applied ($[S_\mathrm{f}^2]$, $[S_\mathrm{f}^2,\tau_\mathrm{f}]$, $[S_\mathrm{s}^2,\tau_\mathrm{s},S_\mathrm{f}^2]$, or $[S_\mathrm{s}^2,\tau_\mathrm{s},S_\mathrm{f}^2,\tau_\mathrm{f}]$ ). Typically, if the model applied varies from one residue to the next, it is accompanied by significant jumps in the model parameters. For example, the $\beta$1-$\beta$2 turn (residues 7-13) exhibits more motion than surrounding residues. We would expect this motion to be partly correlated among these residues, so that the correlation times should be similar. Yet, between residues 9, 10, and 11, $\tau_\mathrm{s}$ varies by about half an order of magnitude, with noticeable variation for the other residues as well (where no strong variation appears in the raw data, Fig. S9). Indeed, three different models were employed to analyze relaxation for β1-β2 turn: a simple model-free model for residue 12, an extended model-free model with no correlation time for fast motion (too fast to be determined) for residues 7, 8, 10 and 13, and a full extended model-free with two defined correlation times for residues 9 and 11.

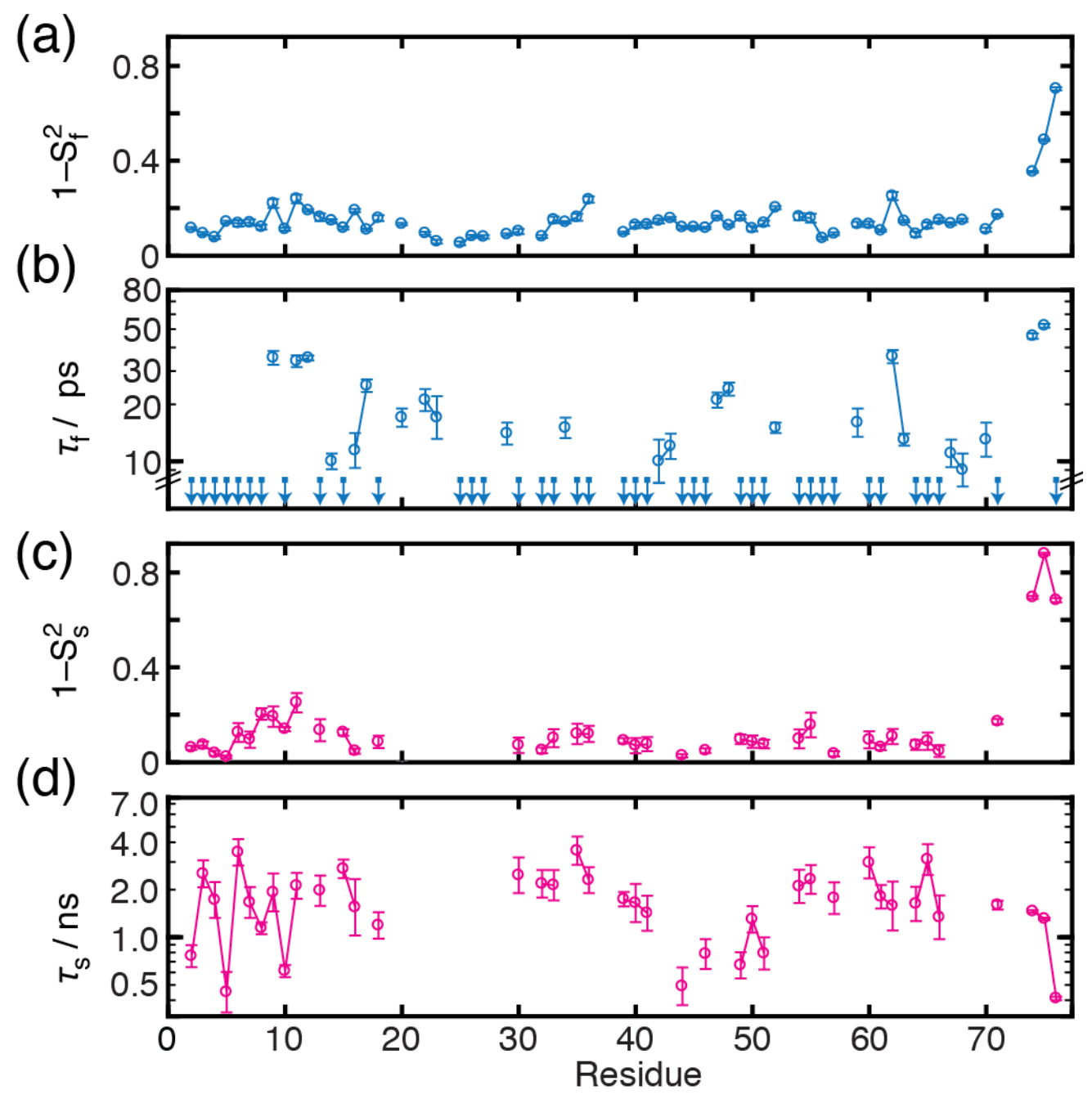

**Fig. 10.** Model-free analysis of ubiquitin high-field data as previously reported by Charlier et al.,[31] using the same data as in Fig. 8(b). (a) shows $(1-S_f^2)$ for the fast motion. (b) plots $\tau_f$, the correlation time of the fast motion. In some cases, $S_f^2$ is fitted but $\tau_f$ is not, where it is assumed the motion is $\tau_f$ is too short to induce relaxation, as indicated with a downward pointing arrow (below 10 ps on the y-axis). (c) and (d) plot slow motion, showing $(1-S_s^2)$ and $\tau_s$ respectively. In some cases, only one motion was fitted, which is then displayed as a fast motion.

The interpretation of these jumps in models and parameters is not trivial: they might be due to real differences in local motions or, perhaps more likely, to small fluctuations of the measured rates or experimental noise that skew the model selection, one way or another. It is thus difficult to interpret all correlation times as true correlation times and it is safer to consider these as effective correlation times, potentially representing multiple motions or motions defined by multiple correlation times. Model selection is considered a necessary evil in order to make the most of the information content of relaxation datasets. Alternatives to model selection have been suggested, where a model-free approach[38] or a different model[4,5] is used consistently to analyze an entire relaxation dataset. In contrast, detectors can be applied without model selection between residues, and when dynamics may be explained simply (fewer parameters), some of the detector responses simply approach zero, without requiring a new model, significantly reducing discontinuity in the resulting parameters.

Variation in model-free analysis parameters may also occur without model selection. The longest stretch of residues analyzed using a single model occurs on residues 2-8 (3 parameters). Significant variation occurs for $\tau_s$, sometimes exceeding an order of magnitude between neighboring residues, and this variation is accompanied by smaller jumps in $(1-S_s^2)$, with longer correlation times correlated with larger values of $(1-S_s^2)$. The strongest outliers for $\tau_s$ are found at residues 2 and 5, where differences are driven by sharp reduction in $R_1$ at these residues (Fig. S9), which can be explained by less motion at long correlation times or more motion at short correlation times (see Fig. 5(b), (d)). In the model-free analysis, the changes in $R_1$ are explained as a decrease in $\tau_s$ and a decrease in $(1-S_s^2)$.

The question arises- is there really a motion at residues 2 and 5 with a shorter correlation time (~$10^{-9.2}$ s = 630 ps) that is absent at the surrounding residues? The detector analysis suggests this does not need to be the case: $\rho_3^{(\theta,S)}$ is particularly sensitive to this range of correlation times, and shows almost no change in detector response. Instead, it explains the decrease in $R_1$ at residue 2 as a decrease in motion at longer correlation times ($\rho_4^{(\theta,S)}$), and the decrease in $R_1$ at residue 5 as an increase in motion at short correlation times ($\rho_1^{(\theta,S)}$). Differences in responses of the two residues results from differences in the experimental $R_2$ data (Fig. S9). Note that this does not *disprove* the results of model-free analysis: it is possible that the simultaneously decreasing values of $\tau_s$ and $(1-S_s^2)$ cancel each other out, resulting in the apparent uniformity of $\rho_3^{(\theta,S)}$ (model-free and detector analyses should usually be consistent, but detectors can be more broadly interpreted[13]). However, just as model selection requires us to consider that the fitted correlation times may be effective correlation times, even if only a single model is applied, we must still consider that the resulting parameters are effective and represent multiple motions (see Fig. 2). When this is the case, detectors give a more direct picture of the distributions of motion that lead to the effective model parameters.

The limitations of the use of a single effective correlation time were discussed in the original article by Lipari and Szabo.[9] In particular, Lipari and Szabo showed that drastically different distributions of correlation times can lead to very similar observables (see Figure 3 of ref. [9]), particularly when the spectral density function is only probed at a handful of frequencies. Thus, one should keep in mind that correlation times in the model-free approach are effective. By contrast, the detector analysis provides information about the amplitude of motion over a given range of frequencies, and have a well-defined relationship to the distribution of motion (Eq. (20)). This information is less model-dependent and less prone to over-interpretation. In addition, the use of a single model to analyze relaxation rate constants for the entire protein makes direct comparison between given residues easier, and facilitates the interpretation of variations of detector responses.

### B. Relationship to the LeMaster approach

The limits of conventional model-free analysis have motivated development of the dynamic detectors method of analysis, and its subsequent adaptation for solution-state dynamics. Other alternative methods have been proposed to analyze relaxation data sets. For example, the spectral-density mapping method[14-16] of analyzing relaxation data acquired at one or several magnetic fields ($R_1$, $R_2$, $\sigma_{\mathrm{NH}}$ rate constants) avoids distortion of dynamic information. It turns out that spectral density mapping is a special case of dynamics detectors (as previously discussed, see ref. [13] section IIID). However, spectral-density mapping only yields the spectral densities at a few frequencies. Thus, it does not provide directly quantitative information about correlation times. In addition, spectral-density mapping describes the total motion including tumbling, forgoing the separation of internal motion and tumbling motion. LeMaster addressed this limitation,[21] by introducing an alternative analysis of solution-state relaxation data acquired at a single field which accounts for tumbling. In his approach, he suggested fitting the three relaxation rate constants to a spectral density of the following form:

$$J(\omega_i)=\frac{2}{5}S_{\mathrm{f}}^2\left[S_{\mathrm{H}}^2S_{\mathrm{N}}^2\frac{\tau_{\mathrm{r}}}{1+(\omega_i\tau_{\mathrm{r}})^2}+(1-S_{\mathrm{H}}^2)\frac{\tau_{\mathrm{H}}}{1+(\omega_i\tau_{\mathrm{H}})^2}+S_{\mathrm{H}}^2(1-S_{\mathrm{N}}^2)\frac{\tau_{\mathrm{N}}}{1+(\omega_i\tau_{\mathrm{N}})^2}\right]. \tag{27}$$

Rather than having five free parameters for each residue ( $S_{\mathrm{f}}^2, S_{\mathrm{H}}^2, S_{\mathrm{N}}^2, \tau_{\mathrm{H}}, \tau_{\mathrm{N}}$ , where $\tau_r$ is determined from the complete data set of all residues), LeMaster proposed fixing $\tau_H=1/(\omega_H+\omega_N)$ and $\tau_N=-1/\omega_N$, so that $(1-S_{\mathrm{f}}^2)$ is the amplitude of motion for short correlation times, $(1-S_{\mathrm{H}}^2)$ characterizes motion for correlation times nearest to $\tau_{\mathrm{H}}$ , and $(1-S_{\mathrm{N}}^2)$ characterizes motion for correlation times nearest to $\tau_{\mathrm{N}}$. This model accounts explicitly for the bias in the frequencies at which the spectral density is probed by relaxation.

If we rearrange the spectral density as follows,

$$S_f^2 S_H^2 S_N^2 = 1-(1-S_f^2 S_H^2 S_N^2) = 1-\left(1-S_f^2+S_f^2(1-S_H^2)+S_f^2 S_H^2(1-S_N^2)\right)$$

$$J(\omega_i)=\frac{2}{5}\left[\frac{\tau_r}{1+(\omega_i\tau_r)^2}+(1-S_f^2)\frac{-\tau_r}{1+(\omega_i\tau_r)^2}+S_f^2(1-S_H^2)\left(\frac{\tau_H}{1+(\omega_i\tau_H)^2}-\frac{\tau_r}{1+(\omega_i\tau_r)^2}\right),\right.$$
$$\left.+S_f^2 S_H^2(1-S_N^2)\left(\frac{\tau_N}{1+(\omega_i\tau_N)^2}-\frac{\tau_r}{1+(\omega_i\tau_r)^2}\right)\right] \quad (28)$$

we see that the spectral density is a linear function of $(1-S_f^2)$, $S_f^2(1-S_H^2)$, and $S_f^2 S_H^2(1-S_N^2)$, with a fixed offset term. In fact, the coefficients of these three terms include a negative contribution due to internal motion attenuating relaxation from tumbling. This is similar to the design of sensitivities to internal motion ($R_\zeta^{\text{solu.}}(z_i)$) in the detector analysis (Eq. (18)).

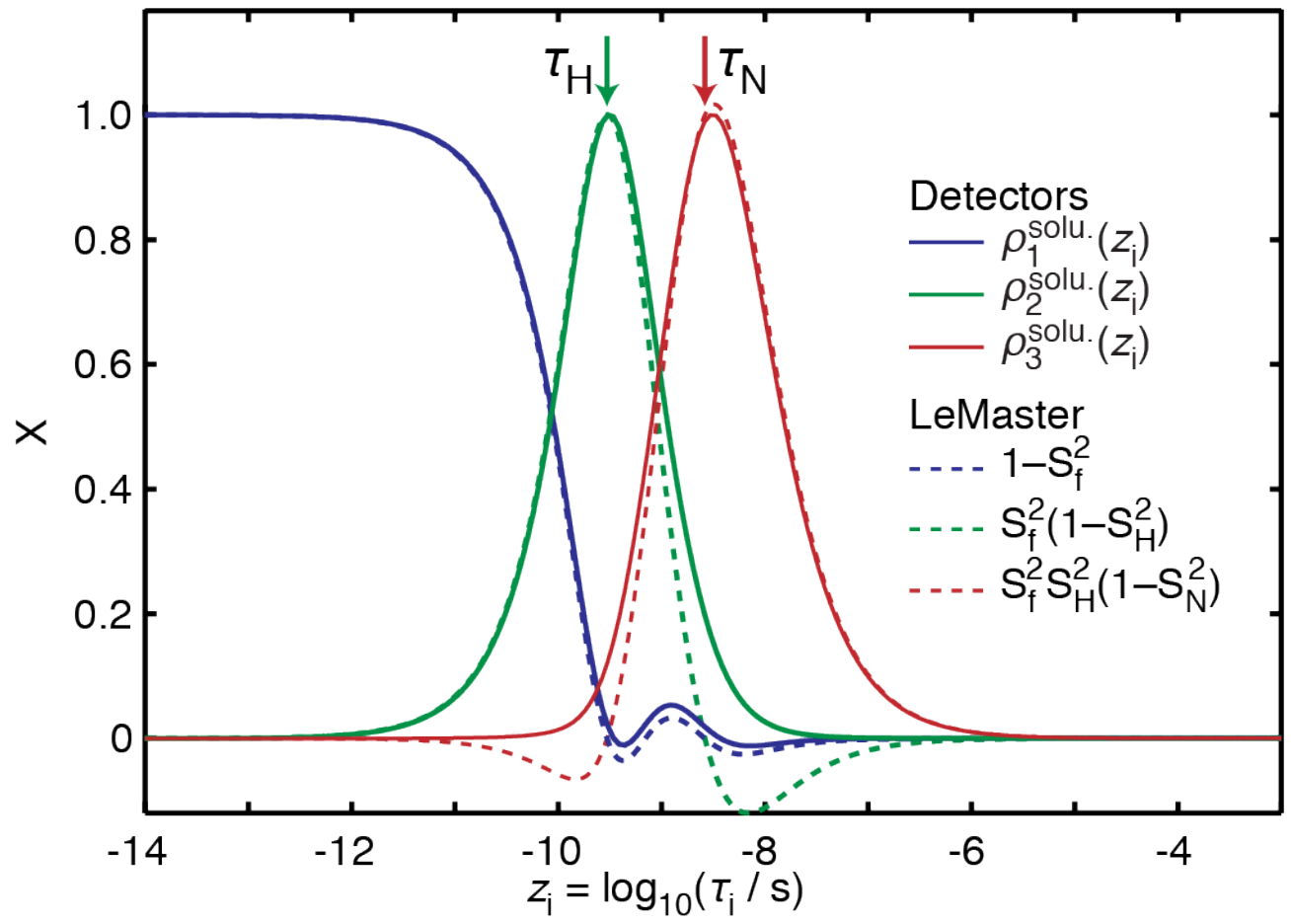

**Fig. 11.** Detectors vs. LeMaster approach. Solid lines show three detector sensitivities are optimized from $R_1$, $R_2$, and $\sigma_{NH}$ rate constants at 600 MHz, assuming a rotational correlation time of 4.84 ns. Dashed lines show the sensitivities of the three terms resulting from the LeMaster approach. Arrows indicate the position of $\tau_H$ and $\tau_N$. One sees that the resulting behavior is very similar, although the LeMaster approach results in more regions of negative sensitivity.

We have derived the sensitivity of these three terms as a function of correlation time and compared them to the detector sensitivities (Fig. 11). Detector sensitivities and amplitudes of the terms in LeMaster's approach (Eq. (28)) are remarkably similar to the detector sensitivities for the same data set. Furthermore, the correlation times for the maximum sensitivities of $\rho_2$ and $\rho_3$, and $S_f^2(1-S_H^2)$ and $S_f^2 S_H^2(1-S_N^2)$, nearly coincide with $\tau_H$ and $\tau_N$, the fixed correlation times used by LeMaster. When comparing to detector sensitivity, one notes that the amplitudes using LeMaster's approach become slightly more negative for some correlation times, which may lead to small differences as compared to the detectors approach. LeMaster's approach is a special case of the more general detectors approach for relaxation datasets recorded at a single magnetic field.

## IV. Conclusions

Dynamics detectors have been developed to characterize distributions of motion of arbitrary complexity using solution-state NMR relaxation data. A set of detectors is optimized for a given experimental relaxation dataset, where each detector characterizes the amount of motion for a well-defined range of correlation times. The approach is an adaptation of the concept developed for solid-state NMR relaxation data. We obtain detectors that are sensitive to the internal motion of a molecule tumbling in solution, but are not sensitive to the tumbling motion itself. This is accomplished by defining rate-constant sensitivities to the internal motion for molecules tumbling isotropically in solution, and obtaining detectors from these sensitivities. Detector analysis does not suffer from the biases of model-free/extended model-free analyses of relaxation data: when using model-free formalism to analyze relaxation data with an underlying complex distribution of motion, the results are difficult to interpret in terms of the physical motion. We apply the detector method to the analysis of $^{15}$N relaxation rate constants in ubiquitin, and find a more easily interpretable and stable description of internal dynamics than is obtained with conventional model-free analysis. This demonstrates the utility of the detector approach in solution-state NMR.

## V. Acknowledgements

This work was supported by the Swiss National Science Foundation (Grants 200020_159707 and 200020_178792). This project has also received funding from the European Research Council (ERC) under the European Union's Horizon 2020 research and innovation program (grant agreement nº 741863, FASTER) and under the European Union's Seventh Framework Programme (FP7/2007-2013), ERC Grant agreement 279519 (2F4BIODYN) (to F.F.).

*Supplementary information for:*

# Reducing bias in the analysis of solution-state NMR data with dynamics detectors

Albert A. Smith[1,2], Matthias Ernst[1], Beat H. Meier[1], Fabien Ferrage[3]

[1]*ETH Zurich, Physical Chemistry, Vladimir-Prelog-Weg 2, 8093 Zurich, Switzerland*

[2]*Present address: Universität Leipzig, Insitut für Medizinische Physik und Biophysik, Härtelstraße 16-18, 04107 Leipzig, Germany*

[3]*Laboratoire des biomolécules, LBM, Département de chimie, École normale supérieure, PSL University, Sorbonne Université, CNRS, 75005 Paris, France.*

M.E. : maer@ethz.ch

B.M. : beme@ethz.ch

F.F. : Fabien.Ferrage@ens.fr

## Table of Contents

## Glossary of terms

| Name | Symbol | Units | Description |
|---|---|---|---|
| *Correlation time* | $\tau_c$ | s | Correlation time of some motion in the system. |
| *Log-correlation time* | $z$ | unitless (vs. 1 s) | Base-10 logarithm of the correlation time, given by $\log_{10}(\tau_c / 1\,\mathrm{s})$. |
| *Rotational diffusion correlation time* | $\tau_r$ | s | Correlation time of isotropic rotational of a molecule in solution (tumbling). |
| *Log-$\tau_r$* | $z_r$ | unitless (vs. 1 s) | Base-10 logarithm of the rotational correlation time, given by $z_r = \log_{10}(\tau_r / 1\,\mathrm{s})$ |
| *Effective correlation time* | $\tau_{eff}$ | s | Effective correlation of an internal motion, where the molecule is undergoing tumbling with correlation time, $\tau_r$. Given by $\tau_{eff} = \tau_r \tau_c \big/ \left(\tau_r + \tau_c\right)$. |
| *Log-effective correlation time* | $z_{eff}$ | unitless (vs. 1 s) | Base-10 logarithm of the effective correlation time, given by $\log_{10}(\tau_{eff} / 1\,\mathrm{s})$. |
| *Distribution of motion* | $(1-S^2)\theta(z)$ | unitless | Describes how motion is distributed as a function of correlation time, where $z = \log_{10}(\tau_c / 1\ \mathrm{s})$. $(1–S^2)$ gives the total amplitude of motion, so that $\theta(z)$ always integrates to one. |
| *Distribution of internal motion* | $(1-S^2)\theta(z_i)$ | unitless | This is the same as the distribution of motion for solid-state analysis. In solution-state analysis, this distribution only accounts for internal motion of the molecule- in other words, tumbling of the molecule is factored out, and the log-correlation times are *not* effective correlation times (see SI section 1 for comparison of distributions). |
| *Distribution of total motion* | $\theta_{\mathrm{tot.}}(z)$ | unitless | This is the distribution of all motions for a molecule tumbling in solution, including the tumbling itself. Motion resulting from internal motion is modified to have an effective correlation time, $z_{eff}$, which results from the internal correlation time and the tumbling correlation time see SI section 1 for comparison of distributions). |
| *Relaxation rate constant* | $R_\zeta^{(\theta,S)}$ | $\mathrm{s}^{-1}$ | The relaxation-rate constant obtained under experimental conditions denoted by $\zeta$, for a distribution of motion $(1–S^2)\theta(z)$. May be obtained by integrating the product of the sensitivity of that rate constant, $R_\zeta(z)$, times the distribution of motion, $(1–S^2)\theta(z)$. |
| *Sensitivity* | $R_\zeta(z)$ | $s^{-1}$ | The relaxation rate constant obtained under experimental conditions denoted by $\zeta$, for a mono-exponential correlation function, having correlation time $\tau_c = 10^z$ s, and amplitude $1–S^2 = 1$. |
| *Solution-state sensitivity* | $R_\zeta^{solu}(z_i)$ | $s^{-1}$ | Sensitivity of an experiment to the internal motion of a molecule, with $\tau_i = 10^{z_i} \cdot 1\,\mathrm{s}$, when the molecule is tumbling in solution. This function has one term to account for attenuation of relaxation due to rotational diffusion, and a second term to account for relaxation induced by the internal motion, given as $R_\zeta^{solu.}(z_i) = R_\zeta(z_{eff}(z_i)) - R_\zeta(z_r)$. |
| *Detector* | – | – | A mathematical tool used to quantify the amount of motion for a range of correlation times. |
| *Detector sensitivity* | $\rho_n(z)$ | unitless | Defines how a detector responds to a particular correlation time, $\tau_c = 10^z$ s. Its value as a function of $z$ is obtained by taking a linear combination of rate constant sensitivities (using the same linear combination as is used to obtain the detector responses). |
| *Detector response* | $\rho_n^{(\theta,S)}$ | unitless | A quantity, describing the amount of motion for a particular range of correlation times, rigorously defined as the integral of the product of the detector sensitivity, $\rho_n(z)$, and the distribution of motion, $(1–S^2)\theta(z)$. Obtained by taking an appropriate linear combination of experimental rate constants (strictly speaking, by fitting a vector of the rate constants to the detection vectors, $\vec{r}_n$). |
| *Normalized rate constant* | $\Re_\zeta^{(\theta,S)}$ | unitless | The relaxation rate constant divided by some normalization constant, $c_\zeta$, to yield a dimensionless relaxation rate constant. For solution state relaxation, we first subtract away the relaxation rate constant obtained for an internally rigid motion, $R_\zeta^0$, such that $\Re_\zeta^{(\theta,S)} = (R_\zeta^{(\theta,S)} - R_\zeta^0)/c_\zeta$. |
| *Allowed region* | – | – | For a given set of experiments, the allowed region is all sets of rate constants ($R_\zeta^{(\theta,S)}$) that can be obtained for any arbitrary distribution of motion, given by $(1–S^2)\theta(z)$. Usually this space is represented in terms of the $\Re_\zeta^{(\theta,S)}$. |
| *Detection vector* | $\vec{r}_n$ | $s^{-1}$ | A vector containing carefully chosen values of the $R_\zeta^{(\theta,S)}$, so that a vector containing the full set of experimentally determined relaxation rate constants is assumed to be a linear combination of all detection vectors, given by $\rho_1^{(\theta,S)}\vec{r}_1 + \rho_2^{(\theta,S)}\vec{r}_2 + \ldots$. |
| *Sum of normalized rate constants* | $\Sigma_\zeta \Re_\zeta^{(\theta,S)}$ | unitless | Sum of all normalized rate constants for an experimental data set, used for calculating the ratio of rates. Note that for the reduced space for internal motion (solution-state), this term is replaced, often by $-\Re_\zeta^{(\theta,S)}$, where the corresponding sensitivity, $R_\zeta(z)$, remains negative for all correlation times (see main text, Eq. (22)). |
| *Ratio of rates* | $\kappa_\zeta$ | unitless | For experimental conditions denoted by $\zeta$, this is the ratio of the normalized rate constants, $\Re_\zeta^{(\theta,S)}$, divided by the sum of normalized rate constants, $\Sigma\Re^{(\theta,S)}$, which is used for defining positions in the reduced space. |
| *Reduced space* | – | – | For a set of experiments, the reduced space is defined by the ratios of rates, $\kappa_\zeta$, for that set of experiments. The dimensionality of this space is one less than the number of experiments- achieved by omitting one of the experiments when calculating the $\kappa_\zeta$. |
| *Reduced vector* | $\vec{\kappa}$ | unitless | Vector of ratios of rates, $\kappa_\zeta$, defining a position in the reduced space. These positions can be used to define detection vectors, although note that the reduced vector only defines the direction of the detection vector, but not the length. |
| *Effective width* | $\Delta z_n$ | unitless (vs. 1 s) | The effective width of a detector is defined as the detector integral divided by its maximum, given on a base-10 log scale. $\Delta z = \int \rho_n(z)dz \Big/ \max(\rho_n(z))$ |
| *Detector center* | $z_n^0$ | unitless (vs. 1 s) | This gives the center of the detector sensitivity, on a logarithmic scale (unitless, with reference to 1 s using a base-10 log). Defined as follows: $z_n^0 = \int z\rho_n(z)dz \Big/ \int \rho_n(z)dz$ |

## 1. Distribution of the total motion vs. distribution of internal motion

In the case of a molecule tumbling isotropically in solution, we may assume that the total correlation is a product of the correlation function of the internal motion and the correlation function of the tumbling, such that

$$\begin{aligned} C(t) &= C_{\mathrm{O}}(t)C_{\mathrm{I}}(t) \\ C_{\mathrm{O}}(t) &= \frac{1}{5}\exp(-t/\tau_{\mathrm{r}}) \end{aligned} \quad \text{(S1)}$$

Here, we write correlation function of the internal motion as

$$C_{\mathrm{I}}(t) = \frac{1}{5}\left[S^2 + (1-S^2)\int_{-\infty}^{\infty}\theta(z_{\mathrm{i}})\exp(-t/(10^{z_{\mathrm{i}}}\cdot 1\,\mathrm{s}))dz\right], \quad \text{(S2)}$$

where $(1\text{–}S^2)\theta(z_i)$ describes the distribution of internal motion (we use $z_i$ to distinguish the correlation time of the internal motion from $z_{eff}$, which appears in the next equation as the effective correlation time). The product of the two correlation functions then yields

$$\begin{aligned} C(t) &= \frac{1}{5}\left[S^2\exp(-t/\tau_{\mathrm{r}}) + (1-S^2)\int_{-\infty}^{\infty}\theta(z_{\mathrm{i}})\exp(-t/(10^{z_{\mathrm{eff}}(z_{\mathrm{i}})}\cdot 1\,\mathrm{s}))dz_{\mathrm{i}}\right] \\ \tau_{\mathrm{eff}}(z_{\mathrm{i}}) &= \frac{(10^{z_{\mathrm{i}}}\cdot 1\,\mathrm{s})\tau_{\mathrm{r}}}{(10^{z_{\mathrm{i}}}\cdot 1\,\mathrm{s})+\tau_{\mathrm{r}}},\quad z_{\mathrm{eff}}(z_{\mathrm{i}}) = \log_{10}(\tau_{\mathrm{eff}}(z_{\mathrm{i}})/1\,\mathrm{s}) \end{aligned} \quad \text{(S3)}$$

One notes, however, that this correlation function is still a sum of decaying exponential terms, although with modified correlation times (given by $z_{\mathrm{eff}}(z) = \log_{10}(\tau_{\mathrm{eff}}/1\,\mathrm{s})$). Therefore, we can analyze relaxation arising from such a correlation function using the detector analysis as derived for solid-state NMR,[1] however, we will not characterize the distribution of internal motion, $(1\text{–}S^2)\theta(z_i)$, but rather some distribution of the total motion, $\theta_{tot.}(z)$,such that

$$C(t) = \frac{1}{5}\int_{-\infty}^{\infty}\theta_{\mathrm{tot.}}(z)\exp(-t/(10^{z}\cdot 1\,\mathrm{s}))dz, \quad \text{(S4)}$$

where the correlation functions in Eqs. (S3) and (S4) are equal. Note that due to the tumbling, the total motion is isotropic, so that $(1\text{–}S^2)$ in this case equals 1, and is therefore omitted from Eq. (S4). We will see during the following derivation, that the $z$ appearing in this equation may denote a log-effective correlation time, $z_{eff}$, or log-correlation time for the tumbling, $z_r$, so that we simply denote this variable as $z$.

Then, we would like to know the relationship between $(1\text{–}S^2)\theta(z_i)$ and $\theta_{tot.}(z)$ for a molecule tumbling with correlation time $\tau_r$. We do so by rearrangement of Eqs. (S3) and (S4).

To begin, we define $\theta'_{\text{tot.}}(z)+S^2\delta(z=z_{\text{r}})=\theta_{\text{tot.}}(z)$, where $z_{\text{r}}=\log_{10}(\tau_{\text{r}}/1\,\text{s})$. Inserting into Eq. (S4), we obtain

$$\begin{aligned}C(t)&=\frac{1}{5}\int_{-\infty}^{\infty}\left(\theta'_{\text{tot.}}(z)+S^2\delta(z-z_r)\right)\exp(-t/(10^z\cdot 1\,\text{s}))dz\\&=\frac{1}{5}\left[S^2\exp(-t/\tau_{\text{r}})+\int_{-\infty}^{\infty}\theta'_{\text{tot.}}(z)\exp(-t/(10^z\cdot 1\,\text{s}))dz\right],\end{aligned}\tag{S5}$$

Addition of the $\delta$-function has produced the first term in Eq. (S3), so that by setting Eqs. (S3) and (S5) equal, we may obtain

$$(1-S^2)\int_{-\infty}^{\infty}\theta(z_{\text{i}})\exp(-t/(10^{z_{\text{eff}}(z_{\text{i}})}\cdot 1\,\text{s})dz_{\text{i}}=\int_{-\infty}^{\infty}\theta'_{\text{tot.}}(z)\exp(-t/(10^z\cdot 1\,\text{s}))dz\quad.\tag{S6}$$

Then, we see that the $z$ in the right side of this equation must be equal to $z_{\text{eff}}(z_{\text{i}})$ if the two integrals are equal. Thus, we simply replace all $z$ with $z_{\text{eff}}$ on the right side

$$(1-S^2)\int_{-\infty}^{\infty}\theta(z_{\text{i}})\exp(-t/(10^{z_{\text{eff}}(z_{\text{i}})}\cdot 1\,\text{s})dz_{\text{i}}=\int_{-\infty}^{\infty}\theta'_{tot.}(z_{\text{eff}})\exp(-t/(10^{z_{\text{eff}}}\cdot 1\,\text{s}))dz_{\text{eff}}\quad,\tag{S7}$$

followed by changing the integration variable on the left side to $z_{\text{eff}}$. To do so, we need to obtain $z_{\text{i}}$ and $dz_{\text{i}}$ in terms of $z_{\text{eff}}$, and furthermore adjust the integration bounds.

From the definition of $\tau_{\text{eff}}$ (Eq. (S3)), we start with

$$\begin{aligned}&\tau_{\text{eff}}=10^{z_{\text{eff}}}=\frac{10^{z_{\text{i}}}\cdot\tau_r}{10^{z_{\text{i}}}+\tau_r}=\frac{10^{z_{\text{i}}+z_r}}{10^{z_{\text{i}}}+10^{z_r}}\\&10^{z_{\text{eff}}}(10^{z_{\text{i}}}+10^{z_r})=10^{z_{\text{i}}}\cdot 10^{z_r}\\&10^{z_{\text{i}}}=\frac{10^{z_{\text{eff}}}\cdot 10^{z_r}}{10^{z_r}-10^{z_{\text{eff}}}}\\&z_{\text{i}}=z_{\text{eff}}+z_r-\log_{10}(10^{z_r}-10^{z_{\text{eff}}})\\&dz_{\text{i}}=dz_{\text{eff}}+dz_{\text{eff}}\frac{10^{z_{\text{eff}}}}{10^{z_r}-10^{z_{\text{eff}}}}=dz_{\text{eff}}\frac{10^{z_r}}{10^{z_r}-10^{z_{\text{eff}}}}\end{aligned}\quad,\tag{S8}$$

Next, we find the upper and lower bounds of the integral

$$\begin{aligned}&\text{Lower bound:} && \text{Upper bound:}\\&z_{\text{i}}=-\infty, && z_{\text{i}}=\infty\\&10^{z_{\text{eff}}}=\frac{10^{-\infty+z_{\text{r}}}}{10^{-\infty}+10^{z_{\text{r}}}}=0\;, && 10^{z_{\text{eff}}}=\frac{10^{\infty+z_{\text{r}}}}{10^{\infty}+10^{z_{\text{r}}}}=10^{z_{\text{r}}}\quad,\\&z_{\text{eff}}=-\infty && z_{\text{eff}}=z_{\text{r}}\end{aligned}\tag{S9}$$

Plugging in, we obtain

$$
\begin{aligned}
&(1-S^2)\int_{-\infty}^{z_\mathrm{r}} \theta(z_\mathrm{eff}+z_\mathrm{r}-\log_{10}(10^{z_\mathrm{r}}-10^{z_\mathrm{eff}}))\exp(-t/(10^{z_\mathrm{eff}}\cdot 1\,\mathrm{s}))\frac{10^{z_\mathrm{r}}}{10^{z_\mathrm{r}}-10^{z_\mathrm{eff}}}dz_\mathrm{eff} \\
&=\int_{-\infty}^{\infty}\theta'_\mathrm{tot.}(z_\mathrm{eff})\exp(-t/(10^{z_\mathrm{eff}}\cdot 1\,\mathrm{s}))dz_\mathrm{eff}
\end{aligned}
\quad , \qquad \text{(S10)}
$$

We see that we may satisfy the equality with the following definition for $\theta'_\mathrm{tot.}(z)$:

$$
\theta'_\mathrm{tot.}(z_\mathrm{eff})=\begin{cases}(1-S^2)\theta(z_\mathrm{eff}+z_\mathrm{r}-\log_{10}(10^{z_\mathrm{r}}-10^{z_\mathrm{eff}}))\dfrac{10^{z_\mathrm{r}}}{10^{z_\mathrm{r}}-10^{z_\mathrm{eff}}}, & z_\mathrm{eff}<z_\mathrm{r} \\ 0 & z\geq z_\mathrm{r}\end{cases} \quad , \qquad \text{(S11)}
$$

Finally, we may calculate $\theta_\mathrm{tot.}(z)$:

$$
\theta_\mathrm{tot.}(z)=\begin{cases}(1-S^2)\theta(z+z_\mathrm{r}-\log_{10}(10^{z_\mathrm{r}}-10^{z}))\dfrac{10^{z_\mathrm{r}}}{10^{z_\mathrm{r}}-10^{z}}, & z<z_\mathrm{r} \\ S^2\delta(z-z_\mathrm{r}) & z=z_\mathrm{r} \\ 0 & z>z_\mathrm{r}\end{cases} \quad , \qquad \text{(S12)}
$$

Note that for the case $z$=$z_r$, this is no longer an effective correlation time, but simply the log-correlation time of the tumbling, so that we define this function in terms of a general log-correlation time, $z$, as opposed to $z_\mathrm{eff}$.

One sees that the result is reasonable. A δ-function introduces the relaxation due to tumbling into the total distribution, so the integral of this term results in the correct amplitude, $S^2$. The effective correlation time cannot exceed the correlation time of the tumbling, so the total distribution becomes zero for $z>z_r$. At very short correlation times, the total distribution becomes equal to the distribution of internal motion ($\log_{10}(10^{z_\mathrm{r}}-10^{z})=z_\mathrm{r}$, $10^{z_\mathrm{r}}/(10^{z_\mathrm{r}}-10^{z})=1$). As the correlation time of the total distribution approaches the rotational correlation time, one uses the effective correlation time in the distribution of internal motion, and further scales up the distribution, since one integrates over a narrower range of correlation times. Fig. S1 illustrates this for two distributions:

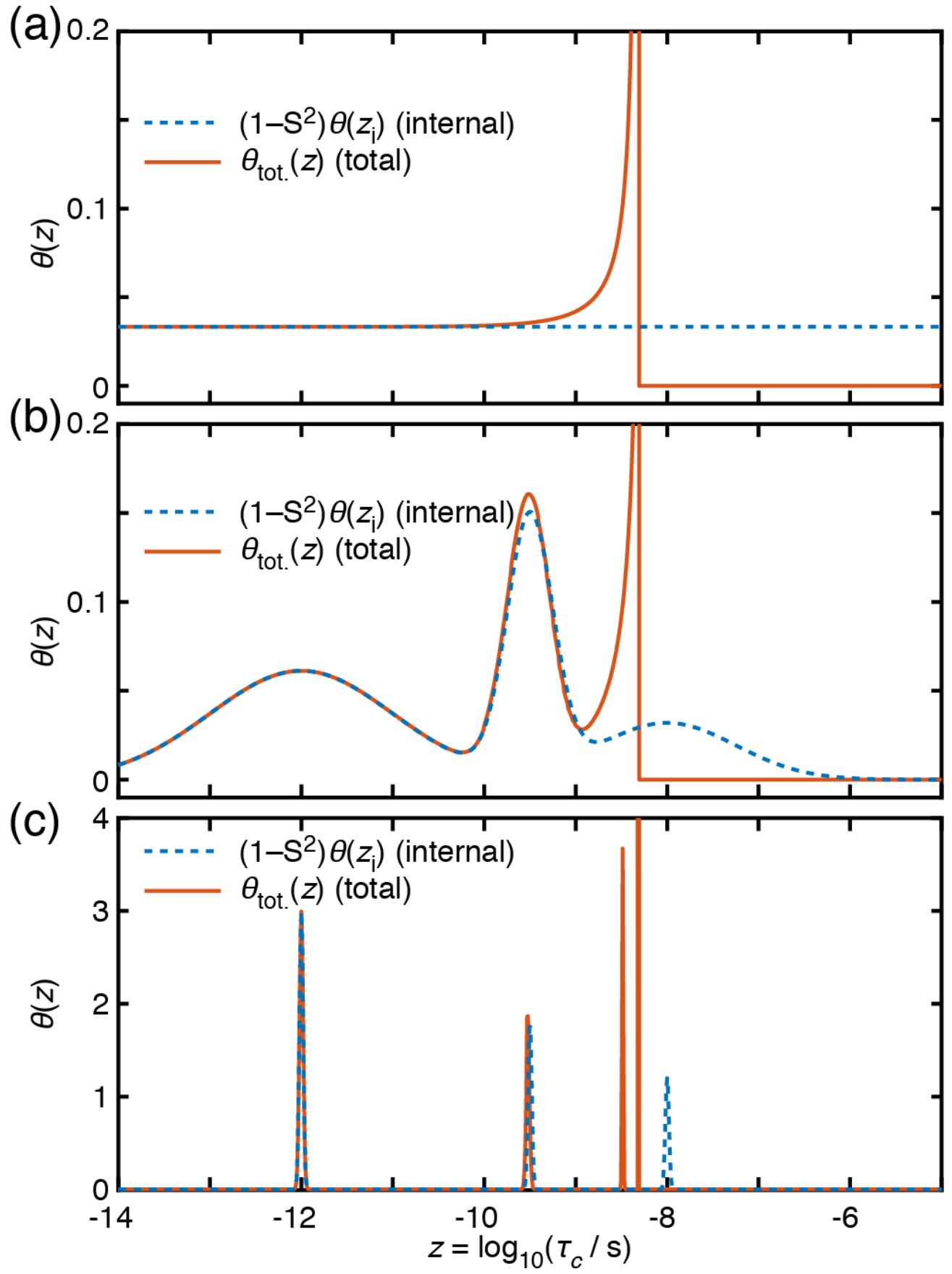

**Fig. S1.** Distributions of internal motion vs. distribution of total motion. Subplots (a)–(c) each show a distribution of the internal motion ($(1–S^2)\theta(z_i)$: blue, dashed line) and the resulting total distribution of motion ($\theta_{tot.}(z)$: red, solid line), assuming a rotational correlation time of $\tau_r = 4.84$ ns. (a) shows a uniform distribution for the internal motion, (b) a distribution resulting from three log-Gaussian distributions for the internal motion, and (c) shows three narrow distributions for the internal motion. Note that at the rotational correlation time, the distribution of total motion diverges to infinity (δ-function), and then falls to zero for all $z>z_r$.

## 2. Singular-value decomposition approach to detector optimization

### 2.1. *Designing the detectors*

For large data sets, the 'spaces' method of detector optimization recently developed becomes increasingly challenging, although is nonetheless very powerful for visualization of the information content of relaxation data.[1] Therefore, we introduce an alternative approach here, which utilizes reduced singular-value decomposition (SVD).[2] We begin with a matrix, **M**, which contains the normalized rate constants for a range of correlation times, for example

$$\mathbf{M}=\begin{bmatrix} \Re_\zeta(z_1) & \Re_\zeta(z_2) & \cdots & \Re_\zeta(z_n) \\ \Re_\psi(z_1) & \Re_\psi(z_2) & \cdots & \Re_\psi(z_v) \\ \vdots & \vdots & \ddots & \vdots \\ \Re_\xi(z_1) & \Re_\xi(z_1) & \cdots & \Re_\xi(z_1) \end{bmatrix}. \tag{S13}$$

where the $\zeta$, $\varphi$, $\xi$ are different experimental conditions, and the $z_i$ are elements of a vector of correlation times (log-spaced over the full range of experiment sensitivity, see Fig. S2(a) for an example). Note that ideally, normalization is done with the standard deviation of that experiment, and in the case of multiple residues, we use the median of the standard deviation (otherwise, we normalize the sensitivity with the maximum of its absolute value).

$$\begin{aligned} \Re_\zeta(z) &= R_\zeta(z)/c_\zeta \\ c_\zeta &= \mathrm{median}(\sigma(R_\zeta)) \end{aligned}. \tag{S14}$$

Then, SVD returns three matrices, such that

$$\mathbf{M}=\mathbf{U}\cdot\Sigma\cdot\mathbf{V}', \tag{S15}$$

where, if **M** is an *m*x*n* matrix, then **U** is an *m*x*m* unitary matrix ($\mathbf{U}^{-1}=\mathbf{U}'$, columns of **U** form an orthonormal basis), **V** is an *n*x*n* unitary matrix, and **Σ** is a diagonal *m*x*n* matrix with non-negative, real numbers on the diagonal. Here, we will typically use the truncated SVD, such that

$$\tilde{\mathbf{M}}=\mathbf{U}_t\cdot\Sigma_t\cdot\mathbf{V}_t', \tag{S16}$$

where $\tilde{\mathbf{M}}$ is the closest approximation to **M,** possible with a matrix of rank *t* ($\mathbf{\Sigma}_t$ contains the *t* largest eigenvalues of **Σ**). Then, $\mathbf{U}_t$ is an *m*x*t* matrix, $\mathbf{\Sigma}_t$ is a *t*x*t* diagonal matrix, and $\mathbf{V}_t'$ is a *t*x*n* matrix.

In principle, we could define the columns of ($\mathbf{U}_t\mathbf{\Sigma}_t$) as our detection vectors (after re-normalization by the $c_\zeta$), and the rows of $\mathbf{V}_t'$ as the corresponding sensitivities. However, we see in Fig. S2(b), that the rows of $\mathbf{V}_t'$ are not well-separated sensitivities. This is straightforward to remedy– we simply take linear combinations of the rows of $\mathbf{V}_t'$ that are optimally separated. An example of such linear combinations is shown in Fig. S2(c).

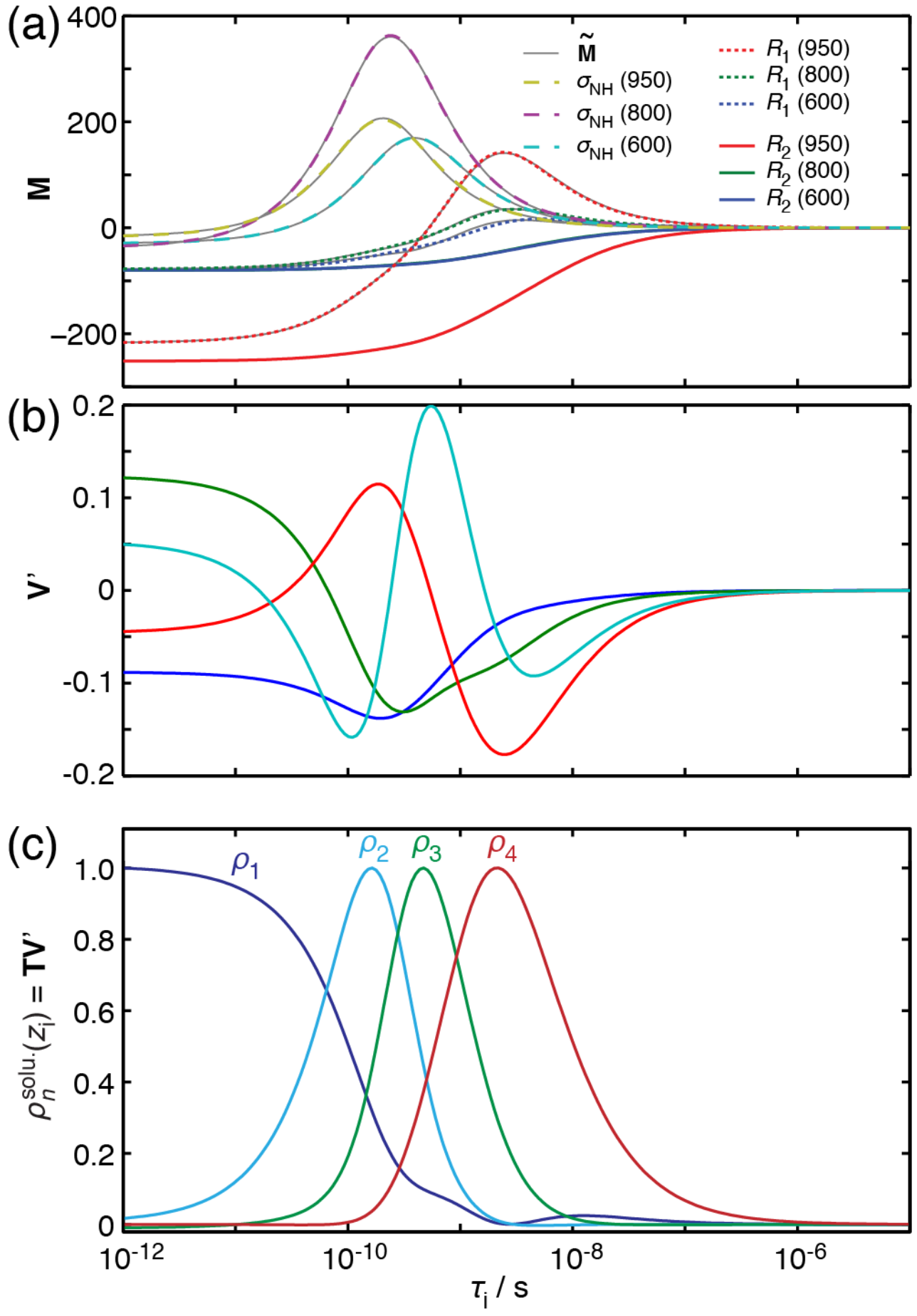

**Fig. S2.** Steps in the singular-value decomposition procedure. Example here are $R_2$, $R_1$, and *NOE* experiments at 600, 800, and 950 MHz. (a) shows rows of the matrix, **M**, which are the sensitivities of the different experiments for a range of correlation times, here normalized by the median standard deviation of that experiment type (taken from [3]). Note that $\tilde{\mathbf{M}}$ is also shown as grey lines (Eq. (S16), for truncated SVD of rank 4), although strongly overlaps with **M** so it is not always visible. (b) shows the rows of **V**', for a truncated SVD of rank 4. (c) shows the detector sensitivities, obtained from linear combinations of the rows of **V**', given by **TV**'.

We denote this transformation as

$$\rho_n(z_m) = \sum_{i=1}^{t} [\mathbf{T}]_{n,i} [\mathbf{V}]_{i,m} , \tag{S17}$$

where **T** is a transformation matrix, for which each row defines a linear combination of the rows in $\mathbf{V}_t$' to yield one of the detector sensitivities, $\rho_n(z)$. Assuming that the $t$ rows of **T** are linearly independent, then $\mathbf{T}^{-1}$ is well-defined, so that one obtains

$$\tilde{\mathbf{M}} = \mathbf{U}_t \cdot \mathbf{\Sigma}_t \cdot \mathbf{T}^{-1} \cdot \mathbf{T} \cdot \mathbf{V}_t' . \tag{S18}$$

If we renormalize $\tilde{\mathbf{M}}$, by multiplying by a diagonal matrix, **c**, which has the normalization constants, $c_\zeta$, along its diagonal we can obtain a matrix that contains the detection vectors along its columns, here referred to as **r**.

$$\mathbf{r} = \mathbf{c} \cdot \mathbf{U}_t \cdot \mathbf{\Sigma}_t \cdot \mathbf{T}^{-1}. \tag{S19}$$

Then, the vector of experimental rate constants is fitted to

$$\min\sum_{\zeta}\sum_{n}\frac{\left(R_{\zeta}^{exp.}-[\mathbf{r}]_{\zeta,n}\rho_{n}^{(\theta,S)}\right)^{2}}{\sigma(R_{\zeta})^{2}}. \tag{S20}$$

where the $\rho_n^{(\theta,S)}$ are variable, or, in matrix form, we solve

$$\min\left|\begin{pmatrix} [\mathbf{r}]_{\zeta,1}/\sigma(R_\zeta) & [\mathbf{r}]_{\zeta,2}/\sigma(R_\zeta) & \cdots & [\mathbf{r}]_{\zeta,n}/\sigma(R_\zeta) \\ [\mathbf{r}]_{\psi,1}/\sigma(R_\psi) & [\mathbf{r}]_{\psi,2}/\sigma(R_\psi) & \cdots & [\mathbf{r}]_{\psi,n}/\sigma(R_\psi) \\ \vdots & \vdots & \ddots & \vdots \\ [\mathbf{r}]_{\xi,1}/\sigma(R_\varphi) & [\mathbf{r}]_{\xi,2}/\sigma(R_\varphi) & \cdots & [\mathbf{r}]_{\xi,n}/\sigma(R_\varphi) \end{pmatrix}\cdot\begin{pmatrix}\rho_1^{(\theta,S)} \\ \rho_2^{(\theta,S)} \\ \vdots \\ \rho_n^{(\theta,S)}\end{pmatrix}-\begin{pmatrix}R_\zeta^{exp.}/\sigma(R_\zeta) \\ R_\psi^{exp.}/\sigma(R_\psi) \\ \vdots \\ R_\xi^{exp.}/\sigma(R_\xi)\end{pmatrix}\right|^2 \tag{S21}$$

where the $|\ldots|^2$ indicates the 2-norm. Note that we restrict the $\rho_n^{(\theta,S)}$ such that $\min\rho_n(z)\leq\rho_n^{(\theta,S)}\leq\max\rho_n(z)$ when solving.

We still must optimize **T**, to give well-separated detector sensitivities. We do so by choosing a target function for each detector ($\rho_n^{target}(z_m)$), and minimizing

$$\sum_m\left|\left(\sum_{i=1}^{t}[\mathbf{T}]_{n,i}[\mathbf{V}]_{i,m}\right)-\rho_n^{target}(z_m)\right|^2, \tag{S22}$$

This has been implemented in the DIFRATE software,[4] as an interactive program with several options for the target function ('SVD_inter.m'), or as a command-line function that takes any user-defined target function ('SVD_target.m').

## 2.2. *Standard deviation of detectors determined from the singular values*

We can estimate the standard deviation of each detector for a given data set, using the singular values. One notes that, if we neglect the requirement that $\min\rho_n(z)\leq\rho_n^{(\theta,S)}\leq\max\rho_n(z)$ then the solution to Eq. (S21) is given by

$$\begin{pmatrix}\rho_1^{(\theta,S)} \\ \rho_2^{(\theta,S)} \\ \vdots \\ \rho_n^{(\theta,S)}\end{pmatrix}=\begin{pmatrix} [\mathbf{r}]_{\zeta,1}/\sigma(R_\zeta) & [\mathbf{r}]_{\zeta,2}/\sigma(R_\zeta) & \cdots & [\mathbf{r}]_{\zeta,n}/\sigma(R_\zeta) \\ [\mathbf{r}]_{\psi,1}/\sigma(R_\psi) & [\mathbf{r}]_{\psi,2}/\sigma(R_\psi) & \cdots & [\mathbf{r}]_{\psi,n}/\sigma(R_\psi) \\ \vdots & \vdots & \ddots & \vdots \\ [\mathbf{r}]_{\varphi,1}/\sigma(R_\varphi) & [\mathbf{r}]_{\varphi,2}/\sigma(R_\varphi) & \cdots & [\mathbf{r}]_{\varphi,n}/\sigma(R_\varphi) \end{pmatrix}^{-1}\cdot\begin{pmatrix}R_\zeta^{exp.}/\sigma(R_\zeta) \\ R_\psi^{exp.}/\sigma(R_\psi) \\ \vdots \\ R_\varphi^{exp.}/\sigma(R_\varphi)\end{pmatrix}. \tag{S23}$$

Since this results in a simple linear combination of the experimental rate constants, we can use the usual propagation-of-error rules to obtain the standard deviation of the detectors (if $r = ax + by + cz$ , then $\sigma^2(r) = a^2\sigma^2(x) + b^2\sigma^2(y) + c^2\sigma^2(z)$ assuming zero covariance). If we take **M** to be the inverse matrix in SI Eq. (S23), then variances for each detector are given by

$$\sigma^2(\rho_n) = \sum_{\zeta} (\mathbf{M}_{n,\zeta})^2 (1)^2 . \tag{S24}$$

The experiments are already normalized by their own standard deviations, so that the variance contribution from each experiment to the detector is just $(1)^2$. Then the variance for each detector is simply the sum of the elements in the corresponding row of the inverse matrix (sum over all experiments, $\zeta$). If we substitute the $c_\zeta$ for the $\sigma(R_\zeta)$ in this matrix (the $c_\zeta$ are just the median of the residue specific $\sigma(R_\zeta)$, so this will change the result slightly, but is a good way to understand the general behavior- see SI Eq. (S19)), the matrix inverse is given by

$$\mathbf{M} = \left(\mathbf{U}_t \cdot \mathbf{\Sigma}_t \cdot \mathbf{T}^{-1}\right)^{-1} = \mathbf{T} \cdot \mathbf{\Sigma}_t^{-1} \mathbf{U}_t' . \tag{S25}$$

Then, the variance for a given detector is given by the 2-norm of the corresponding row of this matrix:

$$\sigma^2(\rho_n^{(\theta,S)}) = \sum_{\zeta} \left[ \mathbf{T}\Sigma_t^{-1}\mathbf{U}_t' \right]_{n,\zeta}^2 \tag{S26}$$

We may simplify this equation by first separating the matrix product into two parts ($\mathbf{T}\Sigma_t^{-1}$ and $\mathbf{U}_t'$ ), inserting a sum over the $t$ singular values, and multiplying out $\mathbf{T}\Sigma_t^{-1}$ (which is straightforward since $\Sigma_t^{-1}$ is diagonal)

$$\begin{aligned} \sigma^2(\rho_n^{(\theta,S)}) &= \sum_{\zeta} \left[ \sum_{i=1}^{t} [\mathbf{T}\Sigma_t^{-1}]_{n,i} [\mathbf{U}_t']_{i,\zeta} \right]^2 \\ \left[\mathbf{T}\Sigma_t^{-1}\right]_{n,i} &= \mathbf{T}_{n,i} [\Sigma_t^{-1}]_{i,i} \\ \sigma^2(\rho_n^{(\theta,S)}) &= \sum_{\zeta} \left[ \sum_{i=1}^{t} \mathbf{T}_{n,i} [\Sigma_t^{-1}]_{i,i} [\mathbf{U}_t']_{i,\zeta} \right]^2 \end{aligned} \tag{S27}$$

We then expand the squared term, to obtain

$$\begin{aligned} \sigma^2(\rho_n^{(\theta,S)}) &= \sum_{\zeta} \left[ \sum_{i=1}^{t} \mathbf{T}_{n,i} [\Sigma_t^{-1}]_{i,i} [\mathbf{U}_t']_{i,\zeta} \right] \left[ \sum_{j=1}^{t} \mathbf{T}_{n,j} [\Sigma_t^{-1}]_{j,j} [\mathbf{U}_t']_{j,\zeta} \right] \\ &= \sum_{i=1}^{t} \sum_{j=1}^{t} \mathbf{T}_{n,i} \mathbf{T}_{n,j} [\Sigma_t^{-1}]_{i,i} [\Sigma_t^{-1}]_{j,j} \sum_{\zeta} [\mathbf{U}_t]_{\zeta,i} [\mathbf{U}_t]_{\zeta,j} \end{aligned} \tag{S28}$$

The rearrangement of the summation order allows us to first sum over the $\zeta$, and because the columns of $\mathbf{U}_t$ are orthonormal, this yields 1 for the inner sum, $\sum_\zeta [\mathbf{U}_t]_{\zeta,i}[\mathbf{U}_t]_{\zeta,j}$, if $i=j$, and 0 otherwise. Therefore, we obtain for the variance

$$\sigma^2(\rho_n^{(\theta,S)}) = \sum_{i=1}^{t}\left(\mathbf{T}_{n,i}[\Sigma_t^{-1}]_{i,i}\right)^2 \tag{S29}$$

Then, the variance of each detector depends on the squared inverse of the singular values with weighting determined by the corresponding row of the **T** matrix. Note that this slightly over-estimates the error, because when actually fitting, one enforces that $\min \rho_n(z) \le \rho_n^{(\theta,S)} \le \max \rho_n(z)$. Without this requirement, experimental noise can push the detector responses outside this range, so that enforcing this requirement removes any such noise that would push the detector responses outside this range.

2.3. *Selecting the number of detectors*

Selecting greater or fewer numbers of detectors has a number of effects. More detectors will yield a better fit of the initial data set. It will also allow one to obtain detector sensitivities covering a narrower range of correlation times. However, inclusion of more detectors also means that one will have smaller singular values in the matrix $\Sigma_t$, which we can see in Eq. (S29), yields higher error for the detector responses because of the inclusion of inverse of the singular values, $[\Sigma]_{i,i}^{-1}$. For an example, we take $R_1$ and NOE rate constants at 600, 800, and 950 MHz, and $R_2$ rate constants at 950 MHz, assuming a rotational correlation time of $\tau_r$=4.84 ns. We then calculate, for different numbers of detectors, the quality of fit of each rate constant vs. correlation time, an optimized set of detectors, and the resulting standard deviation of each detector (we will assume that the standard deviation of each measurement 5% of the maximum of the absolute value of the sensitivity). One sees that, in this case, the fit converges when using ~4 detectors, whereas using more detectors yields negligible improvement in the fit, and the standard deviation for each detector grows significantly.

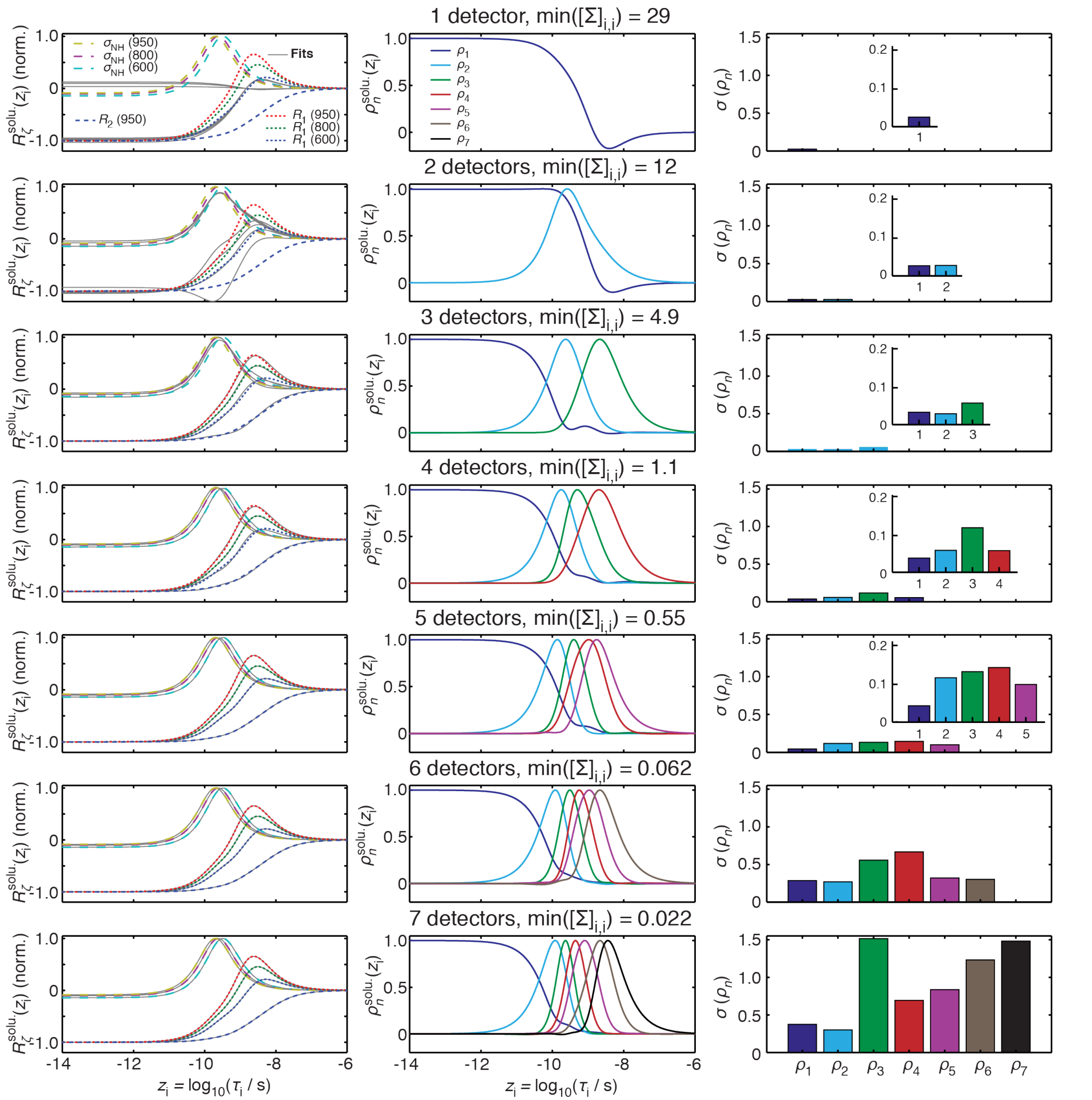

**Fig. S3.** Selection of the number of detectors. Each row uses the number of detectors indicated above the row to fit the set of experiment rate constants. The left column shows the sensitivities of 3 $R_1$ rate constants (600, 800, 950 MHz), 3 NOE rate constants (600, 800, 950 MHz), and 1 $R_2$ rate constant (950 MHz) as colored, dashed lines. Fits of those rate constants using the indicated number of detectors, is shown as solid, grey lines. The middle column shows an optimized set of detectors. The right column shows the standard deviation of each detector, assuming a standard deviation for each rate constant that is 5% of the maximum of the absolute value of the rate constant sensitivity. Inset on some plots shows the same information, scaled up for visibility.

## 3. Ubiquitin analysis at 2 fields

We calculated Ubiquitin dynamics analysis for a data at two fields (600 and 800 MHz, with $R_1$, $R_2$, and NOE data). This is analyzed with 4 detectors, assuming an overall rotational correlation time, $\tau_r$, of 4.84 ns. The results are shown in Fig. S4. Note that we do not treat exchange in this example (compare to main text Fig. 9)

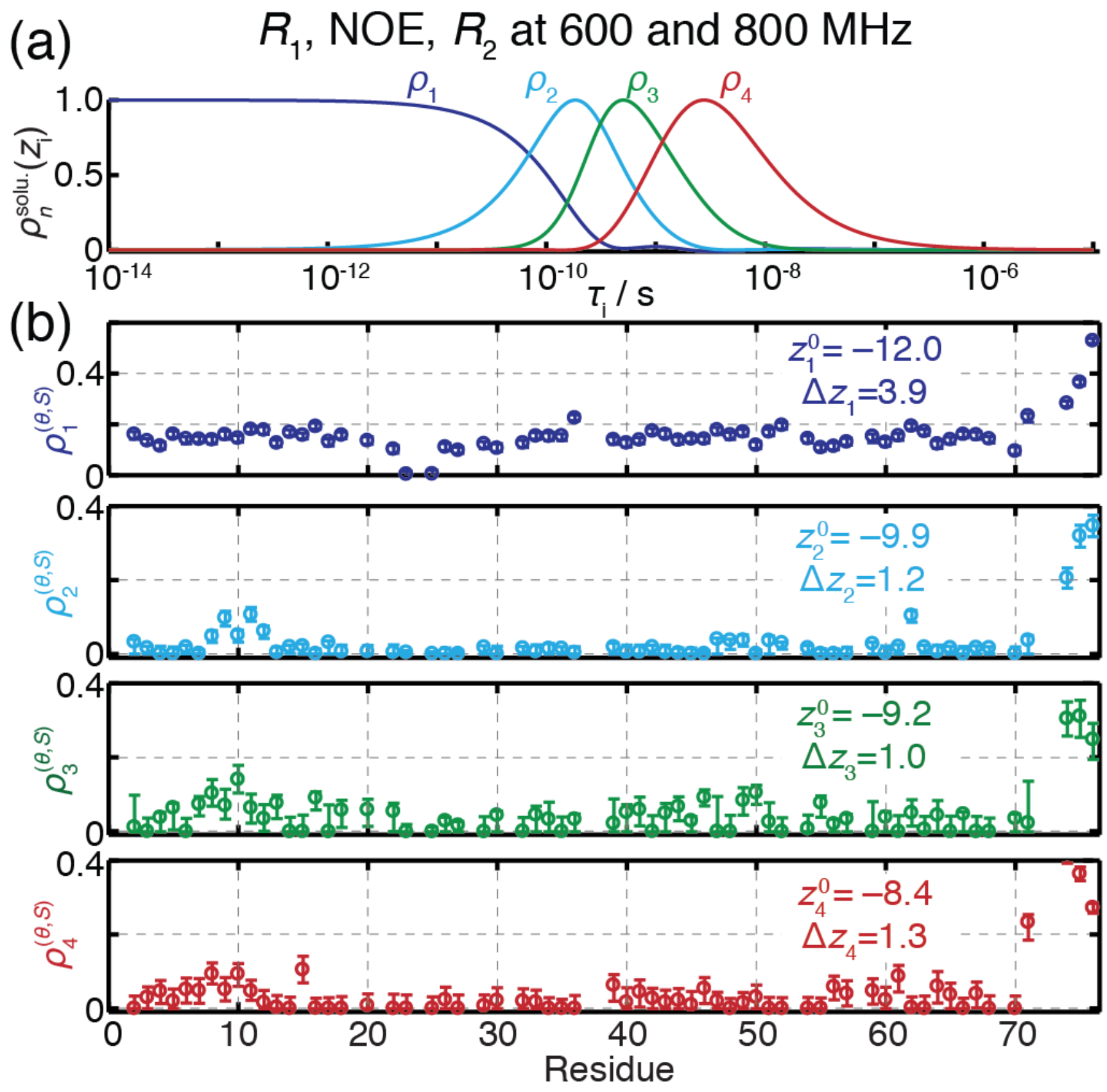

**Fig. S4.** Ubiquitin detector analysis using two fields ($R_1$, $R_2$, NOE at 600 and 800 MHz fields). (a) shows the detector sensitivities, (b) gives the residue-specific detector responses for each of the four detectors. Data fit is shown in Fig. S8, and detection vectors used are given in SI Table S3.

## 4. Model selection for dynamics detectors

In SI section 2.3, the effect of the number of detectors used on fitting and error is discussed. However, this does not tell one how many detectors is best to use. Therefore, we to try to verify that the chosen number of detectors for modeling a particular data set is the best choice, we utilize statistical model selection, via the Akaike Information Criterion (AIC), as well as several variants of this statistical test.[5] The AIC parameter is defined as

$$AIC = N\ln(\chi^2 / N) + 2K\,. \tag{S30}$$

where $\chi^2$ is given by

$$\chi^2 = \sum_{i=1}^{N} \frac{\left(R^i_{exper.} - R^i_{calc.}\right)^2}{\sigma_i^2}, \tag{S31}$$

and $N$ is the total number of experiments, and here $K$ is the number of detectors. Model selection is performed by calculating the AIC parameter and selecting the model with the smallest value.

The AIC assumes a large number of experiments so that it may be biased except in the case that $N>>K$, which is clearly not the case for NMR relaxation studies, possibly resulting in selecting a model that has too many parameters. To counter this, one may use the corrected AIC parameter (AICc),[6,7] defined as

$$AICc = N\ln(\chi^2/N) + 2K + 2\frac{K(K+1)}{(N-K-1)}, \tag{S32}$$

but the correction term is nonetheless not always correct in the case that restrictions are placed on the fitting parameters,[8] as we do when requiring non-negative values for the detector responses. Therefore, we additionally test corrections to the AIC obtained via bootstrapping of the fit.[9] In particular, we use the AICb1 and AICb2 developed by Shang and Cavanaugh,[10] and the 632BQCV statistic developed by Bayer and Cribari-Neto.[8] We calculate the variants of the AIC parameter using data for Ubiquitin acquired at three fields (detector analysis with four detectors found in main text Fig. 9(d)). The results of the AIC tests are shown in Fig. S5.

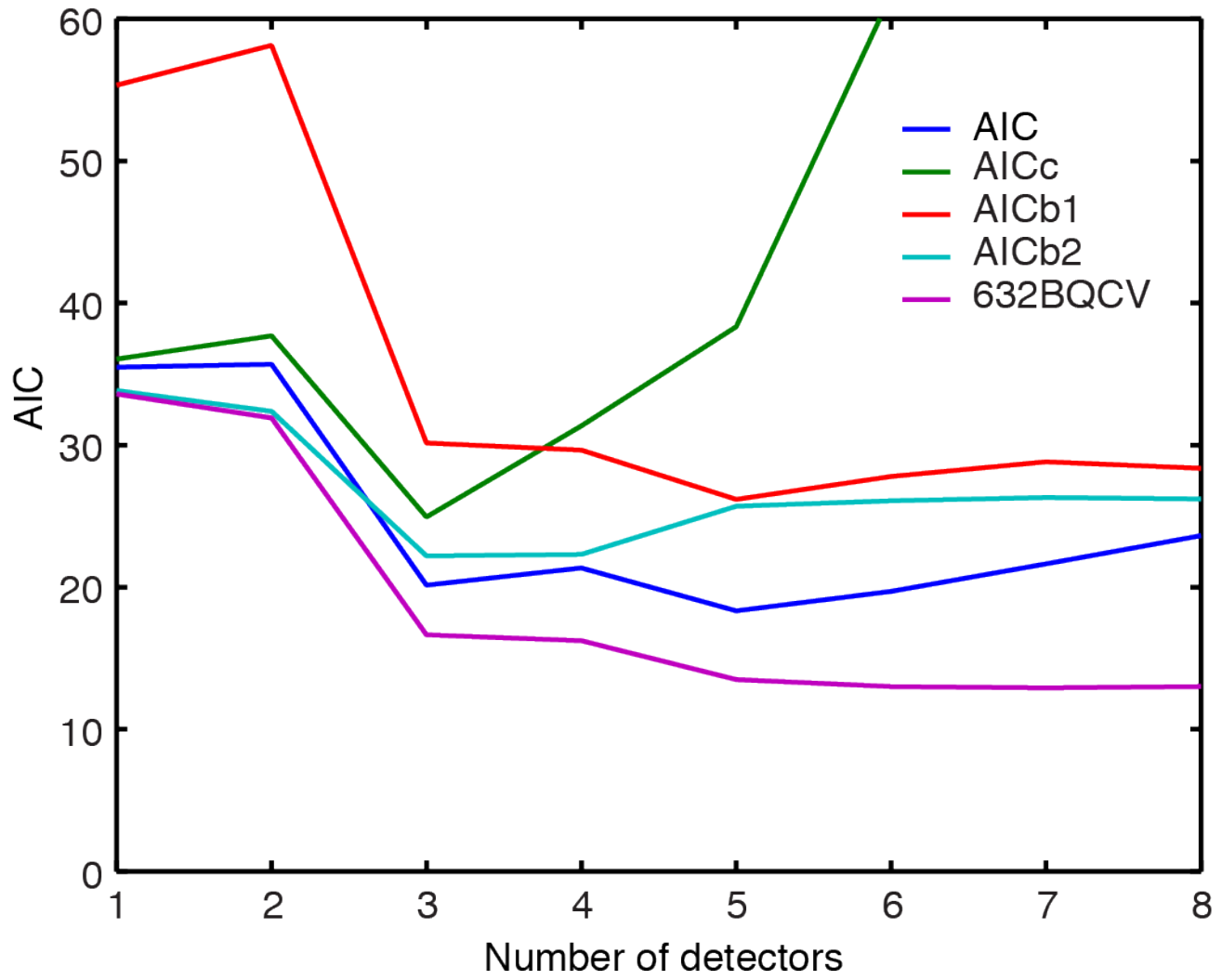

**Fig. S5.** Various AIC parameters as a function of the number of detectors for the analysis of backbone H–N motion in Ubiquitin. A 3 field (600, 800, 950 MHz) data set with $R_1$, $R_2$, and $\sigma_{NH}$ is used. The median AIC value is reported, for all residues.

The AIC selects 5 detectors, whereas the AICc selects only 3 detectors. However, we see that the AICc rises severely for larger number of detectors, in contrast to all other tests. The bootstrap tests (AICb1, AICb2, 632BQCV) do not show a strong preference among models with 4-8 detectors. These tests should be the most reliable since they adapt the AIC correction factor to the specific model behavior based on bootstrap tests. This behavior indicates that as the model increases in complexity, the detector responses contain both more information about the internal motion, but also more noise, such that the models are ultimately of similar quality. Then for detector analysis, one could make the model selection simply based on what one considers an acceptable level of noise on the detector responses. In any case, we see that AICc analysis is not really suitable for model selection for detector analysis, and although the AIC gives similar results to the more rigorous bootstrapped tests, it is not clear that this will always be the case. Then, since AIC takes the assumption that the data set is infinitely large, it is likely better to also avoid this test.

Note that it is not straightforward to obtained a bootstrapped data set from NMR data. Typically, when performing a bootstrap, one takes the original data set, and resamples it randomly, to obtain the bootstrapped data set. However, for relaxation data, this would result in some rate constants being left out entirely, so that our detector responses are not necessarily defined for some possible bootstrapped data sets. This makes this basic approach unfeasible, so that we instead resample the *error* of our fits. Specifically, we take the initial fit to our detectors, back-calculate the rate constants, and calculate the fitting error for each experiment.

$$R^i_{exper.} = R^i_{calc.} + \epsilon_i \,, \tag{S33}$$

Here, $i$ indicates an experiment of the full data set. Then, the bootstrapped data set is given for all rate constants as

$$R^i_{bootstrap.} = R^i_{calc.} + \epsilon_j \frac{\sigma_i}{\sigma_j} \,, \tag{S34}$$

where the index $j$ is selected at random from all experiments in the data set with replacement, and the error is re-scaled according to the standard deviation of the experiment.

## 5. Model-free failure of one- and two-field data sets

As was done with a large relaxation data set in Fig. 1 (main text), it is possible to demonstrate that the model-free approach may not yield a good representation of the true motion in the case of smaller, one- and two-field data sets. We calculate relaxation rate constants here ($R_1$, $R_2$, $\sigma_{NH}$) for motions having three correlation times, such that the correlation function is given by:

$$C(t) = \frac{1}{5}\exp(-t/\tau_r)\left[S^2 + (1-S^2)\sum_{k=1}^{3} A_k \exp(-t/\tau_k)\right], \qquad \text{(S35)}$$

where $\tau_r = 4.84$ ns, and the $A_i$ add to 1. Then, for relaxation rate constants calculated at one field (600 MHz), we fit the data to a correlation function with one internal motion (2 parameter fit):

$$C(t) = \frac{1}{5}\exp(-t/\tau_r)\left[S^2 + (1-S^2)\exp(-t/\tau_1)\right], \qquad \text{(S36)}$$

or two internal motions.

$$C(t) = \frac{1}{5}\exp(-t/\tau_r)\left[S^2 + (1-S^2)(A_1 \exp(-t/\tau_1) + A_2 \exp(-t/\tau_2))\right], \qquad \text{(S37)}$$

Here, we assume the second internal motion is sufficiently fast that it does not directly induce any relaxation, so that its value may be fixed to some arbitrarily small value ($\tau_1 = 10^{-14}$ s, three-parameter fit). Finally, when fitting data with two fields, we use the same correlation function, but allow both correlation times to vary (4 parameter fit).

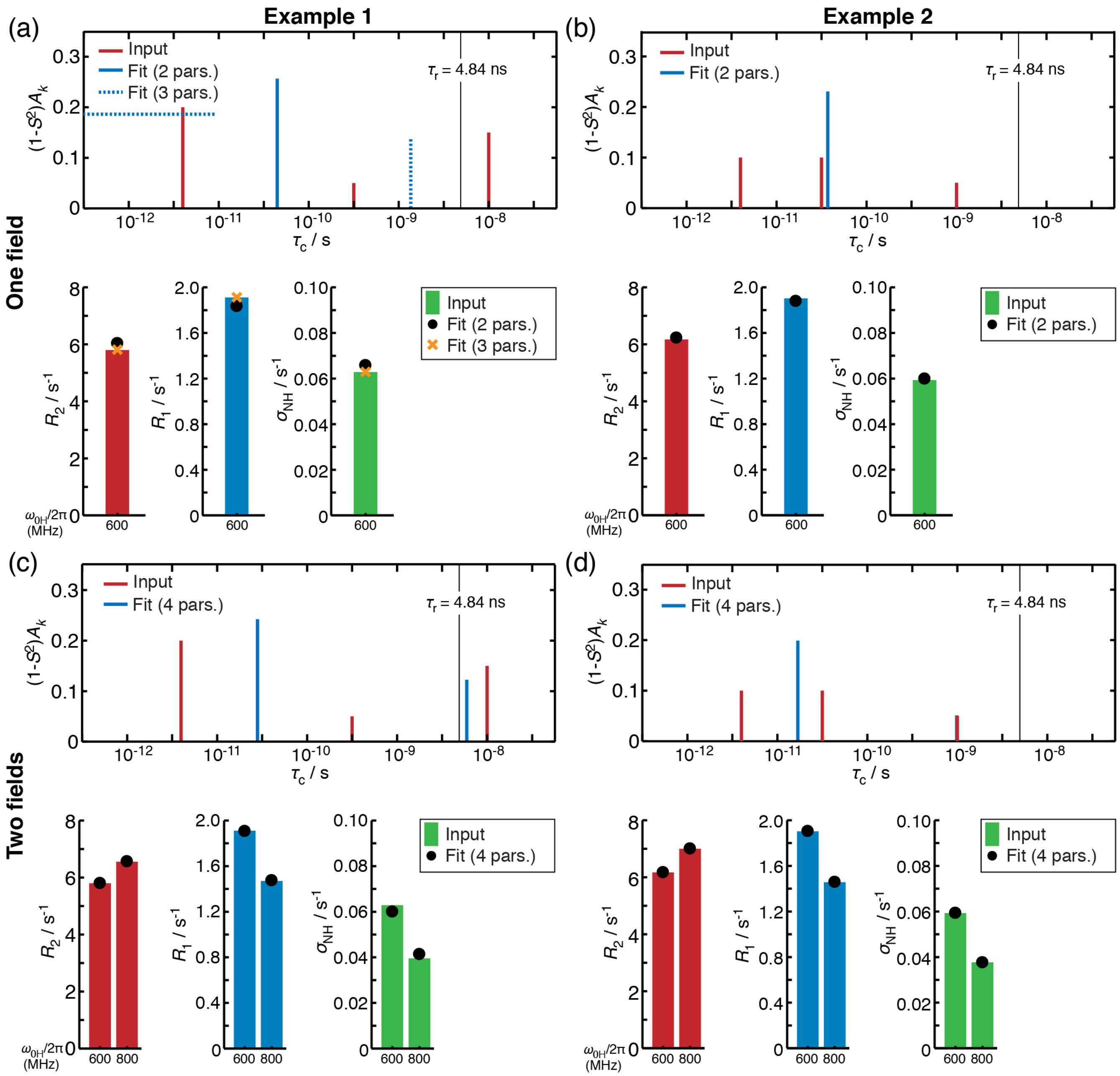

**Fig. S6.** Two correlation functions (example 1: (a), (c) and example 2: (b), (d)) are used to calculate $R_1$, $R_2$, and $\sigma_{NH}$ rate constants at one (600 MHz: (a) and (b)) and two fields (600, 800 MHz: (c) and (d)). The correlation functions are given as line plots in the top of each subplot (red lines, giving the correlation time, $\tau_k$, and amplitude, $(1-S^2)A_k$, of each motion), assuming a rotational correlation time of $\tau_r = 4.84$ ns The resulting rate constants are shown as colored bars in each subplot. These rate constants are then fit to models having 2, 3, or 4 free parameters (see SI Eqs. (S33) and (S34)). The resulting fit parameters are given as blue lines in the top of each subplot, and the fitted rate constants are shown as scatter points in the bottom of each subplot. In (a), the motion is fit both with a two- and three-parameter model. The results for the three-parameter model are shown as dotted lines in the top plot. Since no correlation time is fitted for the faster motion (it is fixed to $\tau_1 = 10^{-14}$ s), it is shown as a horizontal line extending from $\tau_c = 10^{-11}$ s to shorter correlation times.

We see in Fig. S6 that although the data is well-fit in all cases, the fit of the internal motion is usually far away from the input motion, as we expect when the model is simpler than the real motion (note that in Fig. S6(d), the fitted amplitude of the shorter correlation time is approximately the sum of the amplitudes of the two shorter correlation times, and the fitted correlation time converges on the average of these two correlation times, as expected when both motions are in the extreme narrowing limit[11]). Note that in Fig. S6(a), the two parameter yields fitted rate constants that have deviated somewhat from the input, so that we also fit with three parameters, yielding an improved fit of the rate constants.

We may also investigate how well the order parameter of the internal motion is estimated. We tabulate the input and fitted order parameters. We see that the fitted order parameter is always greater than or equal to the fitted order parameter, and note that as the model complexity increases, the accuracy of the order parameter improves (assuming that using a more complex model is justified by poor fit quality of a simpler model). Such a result is expected since tumbling partially or completely masks motions with correlation times comparable to or longer than $\tau_r$. Note that if a motion is not completely masked, then one can improve the estimation of the order parameter by using a more complex model (and including more data in the fit as necessary).

**Table SI.** Input vs. fitted order parameters ($S^2$) for each example and fit.

| | **Input $S^2$ (Ex. 1)** | **Fit $S^2$ (Ex. 1)** | **Input $S^2$ (Ex. 2)** | **Fit $S^2$ (Ex. 2)** |
|---|---|---|---|---|
| **1 field, 2 pars.** | 0.600 | 0.743 | 0.750 | 0.769 |
| **1 fields, 3 pars.** | 0.600 | 0.673 | 0.750 | 0.751* |
| **2 fields, 4 pars.** | 0.600 | 0.638 | 0.750 | 0.750 |

*Fit not shown in Fig. S6

## 6. Plots of data fits

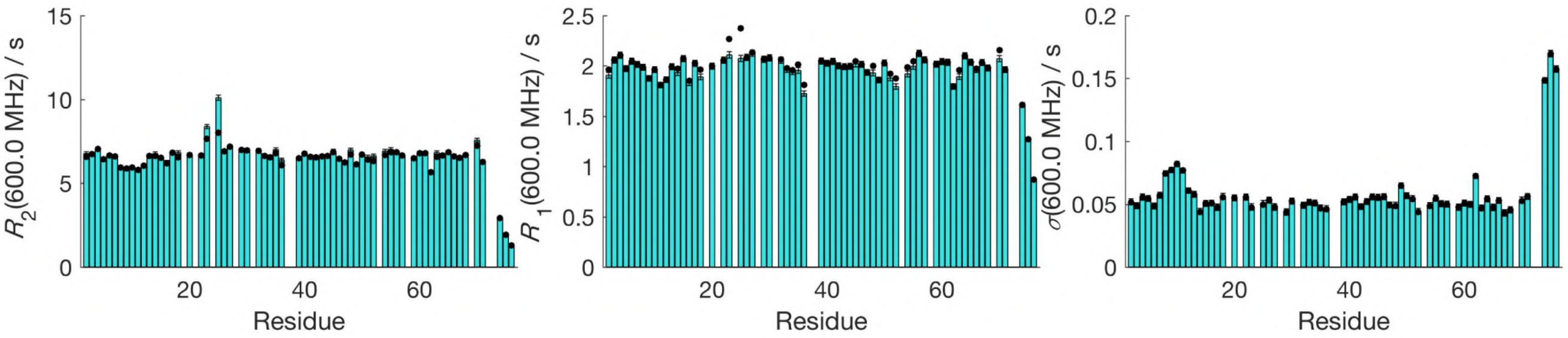

**Fig. S7.** Data fit of ubiquitin using only one $B_0$ field (from analysis shown in Fig. 7B). Each plot shows rate constants for the experiment type shown on the axis (where the field is given in parenthesis). Cyan bars give the value of the rate constant, error bars show one standard deviation, and black circles show the fitted rate constant.

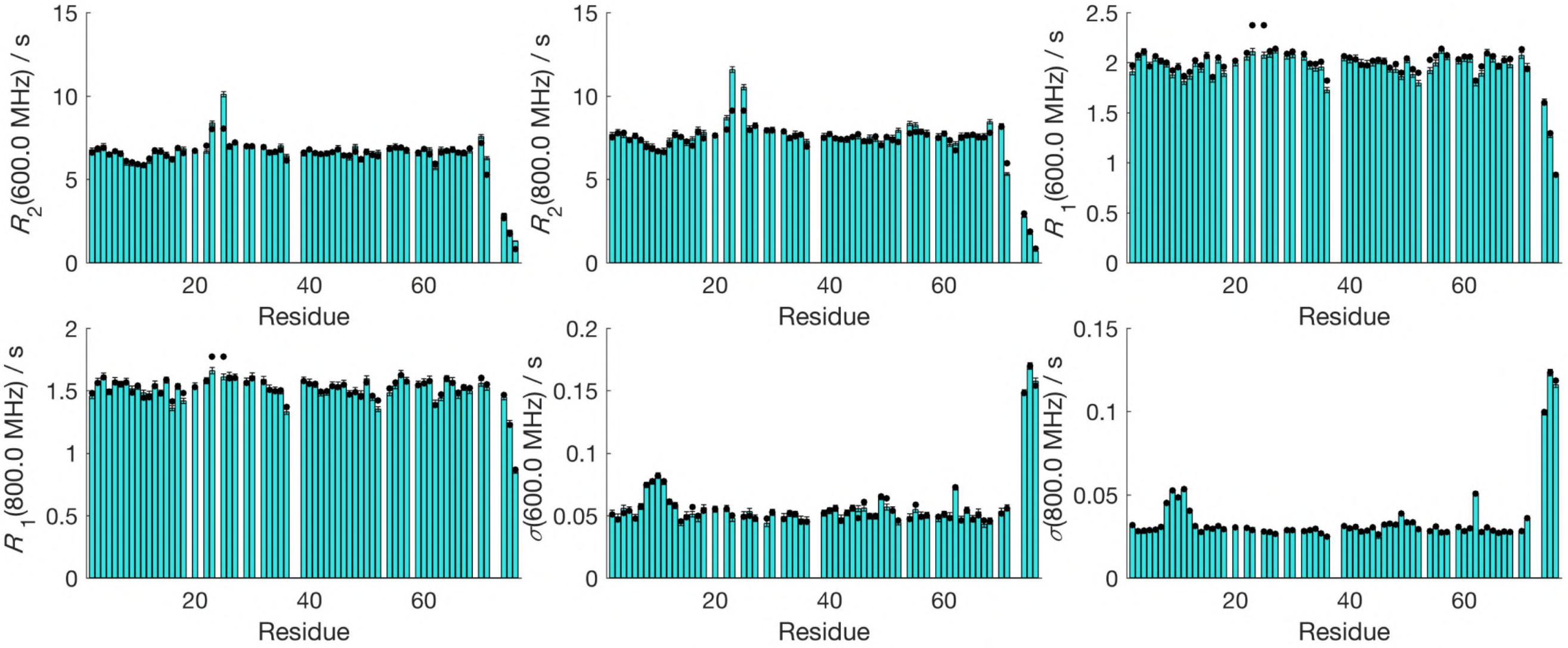

**Fig. S8.** Data fit of ubiquitin using two $B_0$ fields (from analysis shown in SI Fig. S4). Each plot shows rate constants for the experiment type shown on the axis (where the field is given in parenthesis). Cyan bars give the value of the rate constant, error bars show one standard deviation, and black circles show the fitted rate constant.

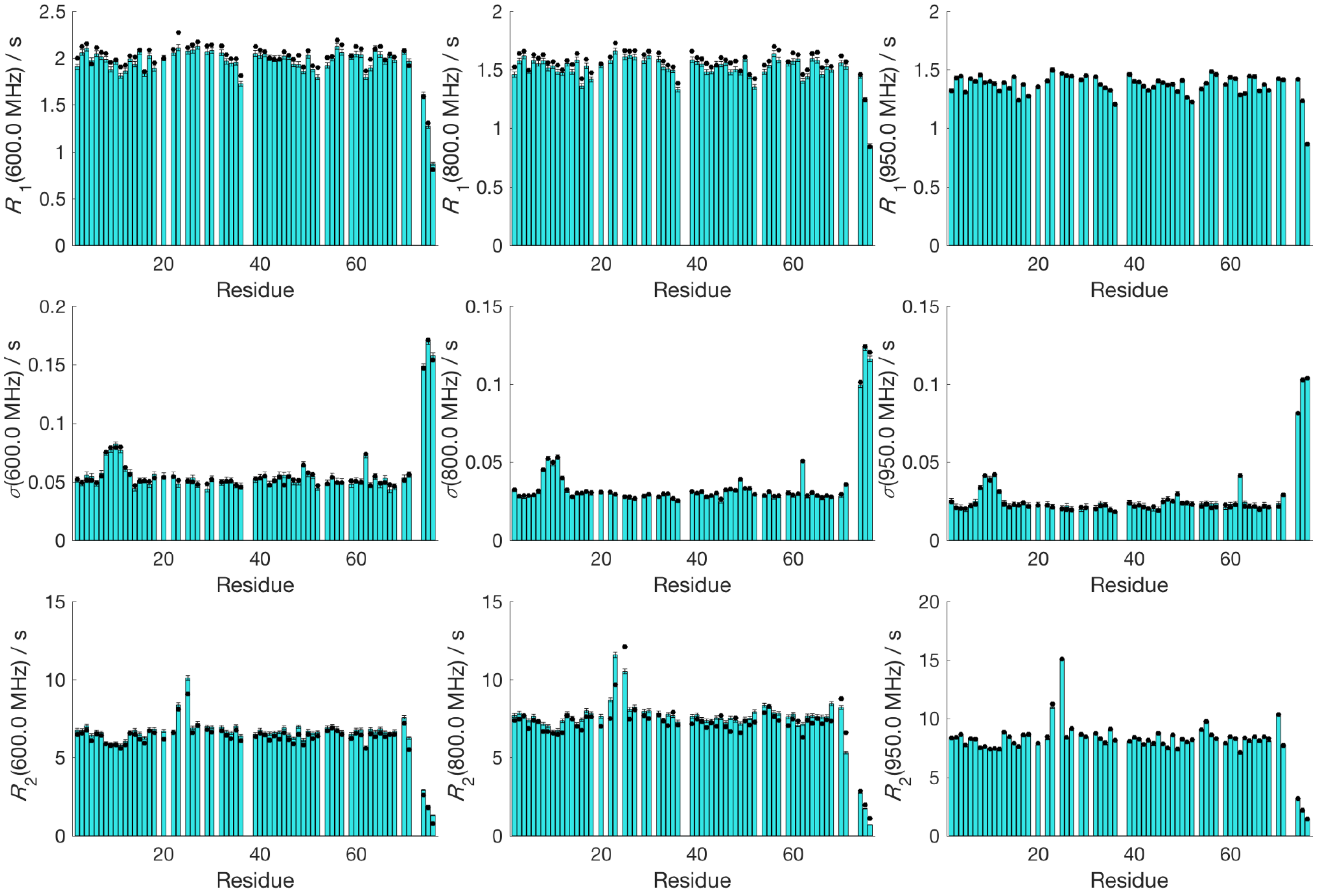

**Fig. S9.** Data fit of ubiquitin using three $B_0$ fields (from analysis shown in Fig. 7(d)). Each plot shows rate constants for the experiment type shown on the axis (where the field is given in parenthesis). Cyan bars give the value of the rate constant, error bars show one standard deviation, and black circles show the fitted rate constant.

# 7. Tables of detection vectors for Ubiquitin analyses

**Table S2: Detection vectors for Ubiquitin analysis at one field (see Fig. 7(a)/(b))**

| | $\vec{r}_1$ / s$^{-1}$ | $\vec{r}_2$ / s$^{-1}$ | $\vec{r}_3$ / s$^{-1}$ | $R_0$ | $B_0$ / T |
|---|---|---|---|---|---|
| $R_{2,600}$ | -8.038 | -6.592 | -2.736 | 8.047 | 14.1 |
| $R_{1,600}$ | -2.381 | -1.347 | 0.662 | 2.382 | 14.1 |
| $\sigma_{HN,600}$ | -0.050 | 0.346 | 0.021 | 0.050 | 14.1 |

*Other parameters:* $\delta_{HN}$=-22945 Hz, $\Delta\sigma_N$=169.5 ppm, $\tau_r$=4.84 ns

**Table S3: Detection vectors for Ubiquitin analysis at two fields (see Fig. S4)**

| | $\vec{r}_1$ / s$^{-1}$ | $\vec{r}_2$ / s$^{-1}$ | $\vec{r}_3$ / s$^{-1}$ | $\vec{r}_4$ / s$^{-1}$ | $R_0$ | $B_0$ / T |
|---|---|---|---|---|---|---|
| $R_{2,600}$ | -8.037 | -3.675 | -3.701 | -2.983 | 8.047 | 14.1 |
| $R_{1,600}$ | -9.123 | -4.271 | -4.213 | -3.542 | 2.382 | 14.1 |
| $\sigma_{HN,600}$ | -2.386 | -0.640 | -0.793 | 0.632 | 0.050 | 14.1 |
| $R_{2,800}$ | -1.776 | -0.390 | -0.322 | 0.874 | 9.140 | 18.8 |
| $R_{1,800}$ | -0.051 | 0.228 | 0.186 | 0.023 | 1.781 | 18.8 |
| $\sigma_{HN,800}$ | -0.028 | 0.234 | 0.078 | 0.019 | 0.028 | 18.8 |

*Other parameters:* $\delta_{HN}$=-22945 Hz, $\Delta\sigma_N$=169.5 ppm, $\tau_r$=4.84 ns

**Table S4: Detection vectors for Ubiquitin analysis at three fields (see Fig. 7(c)/(d))**

| | $\vec{r}_1$ / s$^{-1}$ | $\vec{r}_2$ / s$^{-1}$ | $\vec{r}_3$ / s$^{-1}$ | $\vec{r}_4$ / s$^{-1}$ | $R_0$ | $B_0$ / T |
|---|---|---|---|---|---|---|
| $R_{2,600.3}$ | -8.040 | -2.551 | -3.772 | -3.442 | 8.048 | 14.1 |
| $R_{2,800.4}$ | -9.129 | -2.972 | -4.318 | -4.059 | 9.142 | 18.8 |
| $R_{2,949.4}$ | -10.298 | -3.427 | -4.857 | -4.722 | 10.317 | 22.3 |
| $R_{1,600.3}$ | -2.390 | -0.317 | -0.962 | 0.618 | 2.381 | 14.1 |
| $R_{1,800.4}$ | -1.783 | -0.133 | -0.540 | 0.944 | 1.790 | 18.8 |
| $R_{1,949.4}$ | -1.521 | -0.074 | -0.264 | 1.062 | 1.523 | 22.3 |
| $\sigma_{HN,600.3}$ | -0.050 | 0.182 | 0.254 | 0.038 | 0.050 | 14.1 |
| $\sigma_{HN,800.4}$ | -0.029 | 0.196 | 0.148 | 0.021 | 0.028 | 18.8 |
| $\sigma_{HN,949.4}$ | -0.020 | 0.194 | 0.097 | 0.017 | 0.020 | 22.3 |

*Other parameters:* $\delta_{HN}$=-22945 Hz, $\Delta\sigma_N$=169.5 ppm, $\tau_r$=4.84 ns